\documentclass[11pt]{article}
\makeatletter\renewcommand{\section}{\@startsection
{section}{1}{\z@}{-3.5ex plus -1ex minus
    -.2ex}{2.3ex plus .2ex}{\bf }}

\makeatletter\renewcommand{\subsection}{\@startsection{subsection}{2}{\z@}{
-3.25ex
plus -1ex minus
   -.2ex}{1.5ex plus .2ex}{\it }}
\makeatletter\renewcommand{\subsubsection}{\@startsection{subsubsection}{3}{
-2.45ex}{-3.25ex
plus -1ex minus -.2ex}{1.5ex plus .2ex}{\it }}

\let\fn\footnote
\renewcommand{\footnote}[1]{\linespread{1.1}\fn{#1}\linespread{1.29}}
\usepackage[english]{babel}
\usepackage{mathrsfs}
\usepackage{amsmath,amssymb}
\usepackage{bbm}

\def\slasha#1{\setbox0=\hbox{$#1$}#1\hskip-\wd0\hbox to\wd0{\hss\sl/\/\hss}}

\def\periodb#1{\setbox0=\hbox{$#1$}#1\hskip-\wd0\hbox to\wd0{-}}

\newcommand{\unit}{\mathbbm{1}}   % identity map/matrix
\newcommand{\id}{\mathrm{id}}   % identity map/matrix
\newcommand{\CA}{\mathcal{A}}    % super gauge potential
\newcommand{\CB}{\mathcal{B}}    % super gauge potential
\newcommand{\CF}{\mathcal{F}}    % f-theory
\newcommand{\CG}{\mathcal{G}}    % f-theory
\newcommand{\CI}{\mathcal{I}}    % Lagrangian
\newcommand{\CK}{\mathcal{K}}    % Lagrangian
\newcommand{\CL}{\mathcal{L}}    % Lagrangian
\newcommand{\CCL}{\mathscr{L}}    % real twistor\index{twistor} space
\newcommand{\CCP}{\mathscr{P}}    % real twistor\index{twistor} space
\newcommand{\CCO}{\mathscr{O}}    % real twistor\index{twistor} space
\newcommand{\CN}{\mathcal{N}}    % number of supersymmetries
\newcommand{\CO}{\mathcal{O}}    % bundle classification
\newcommand{\CP}{\mathcal{P}}    % twistor space
\newcommand{\CT}{\mathcal{T}}    % real twistor space
\newcommand{\CU}{\mathcal{U}}    % patches
\newcommand{\CV}{\mathcal{V}}    % Variety
\newcommand{\CE}{\mathcal{E}}    % complex vector bundle
\newcommand{\frH}{\mathfrak{H}}    % complex vector bundle
\newcommand{\frS}{\mathfrak{S}}    % complex vector bundle
\newcommand{\frU}{\mathfrak{U}}    % complex vector bundle
\newcommand{\FR}{\mathbbm{R}}     % field of real numbers
\newcommand{\FC}{\mathbbm{C}}     % field of complex numbers
\newcommand{\CPP}{{\mathbbm{C}P}}    % complex projective plane
\newcommand{\RZ}{\mathbbm{Z}}     % ring of integers
\newcommand{\dd}{\mathrm{d}}     % total differential
\newcommand{\dpar}{\partial}     % partial differential
\newcommand{\dparb}{{\bar{\partial}}}     % partial differential with bar
\newcommand{\embd}{{\hookrightarrow}}     % embedded
\newcommand{\diag}{{\mathrm{diag}}}     % embedded
\newcommand{\di}{\mathrm{i}}     % imaginary unit
\newcommand{\bl}{{\bar{\lambda}}}     % bar \"{u}ber j
\newcommand{\ald}{{\dot{\alpha}}}     % bar \"{u}ber j
\newcommand{\bed}{{\dot{\beta}}}     % bar \"{u}ber j
\newcommand{\gad}{{\dot{\gamma}}}     % bar \"{u}ber j
\newcommand{\ded}{{\dot{\delta}}}     % bar \"{u}ber j
\newcommand{\eps}{{\varepsilon}}     % bar \"{u}ber j
\newcommand{\eand}{{~~~\mbox{and}~~~}}     % bar \"{u}ber j
\newcommand{\der}[1]{\frac{\dpar}{\dpar #1}}   % partielle ableitung
\newcommand{\sU}{\mathsf{U}}     % Euler's number
\newcommand{\sSU}{\mathsf{SU}}     % Euler's number
\newcommand{\sSL}{\mathsf{SL}}     % Euler's number
\newcommand{\sGL}{\mathsf{GL}}     % Euler's number
\newcommand{\sSO}{\mathsf{SO}}     % Euler's number
\newcommand{\sSpin}{\mathsf{Spin}}     % Euler's number
\newcommand{\remark}[1]{}     % Euler's number
\newcommand{\z}[1]{{\stackrel{\circ}{#1}}{}}     % Euler's number
\newcommand{\ed}{{\dot{1}}}    % Functional integral
\newcommand{\zd}{{\dot{2}}}    % Functional integral

\makeatletter \@addtoreset{equation}{section} \makeatother

\begin{document}
\begin{titlepage}
\setcounter{page}{0}
\begin{flushright}
ITP--UH--14/05\\
\end{flushright}
\vskip 1.7cm
\begin{center}
{\LARGE \bf On the Mini-Superambitwistor Space and\\[0.5cm] $\CN=8$ Super
Yang-Mills Theory}
\vskip 1.0cm {\large Christian S\"{a}mann} \vskip 0.8cm
{\em Institut f\"{u}r Theoretische Physik\\
Universit\"{a}t Hannover\\
Appelstra{\ss}e 2, D-30167 Hannover, Germany\\[0.8cm]
{\tt saemann@itp.uni-hannover.de}}\\[5mm]
\vskip 1.5cm
\end{center}
\begin{center}
{\bf Abstract}
\end{center}
\begin{quote}
We construct a new supertwistor space suited for establishing a
Penrose-Ward transform between certain bundles over this space and
solutions to the $\CN=8$ super Yang-Mills equations in three
dimensions. This mini-superambitwistor space is obtained by
dimensional reduction of the superambitwistor space, the standard
superextension of the ambitwistor space. We discuss in detail the
construction of this space and its geometry before presenting the
Penrose-Ward transform. We also comment on a further such
transform for purely bosonic Yang-Mills-Higgs theory in three
dimensions by considering third order formal ``sub-neighborhoods''
of a mini-ambitwistor space.
\end{quote}
\end{titlepage}

\section{Introduction and results}

A convenient way of describing solutions to a wide class of field equations has
been developed using twistor geometry \cite{Penrose:ca,Ward:vs}. In this
picture, solutions to nonlinear field equations are mapped bijectively via the
Penrose-Ward transform to holomorphic structures on vector bundles over an
appropriate twistor space. Such twistor spaces are well known for many theories
including self-dual Yang-Mills (SDYM) theory and its supersymmetric extensions
as well as $\CN$-extended full super Yang-Mills (SYM) theories. In three
dimensions, there are twistor spaces suited for describing the Bogomolny
equations and their supersymmetric variants. The purpose of this paper is to
fill the gaps for three-dimensional $\CN=8$ super Yang-Mills theory as well as
for three-dimensional Yang-Mills-Higgs theory; the cases for intermediate $\CN$
follow trivially. The idea we follow in this paper has partly been presented in
\cite{Popov:2005uv}.

Recall that the supertwistor space describing $\CN=3$ SDYM theory is the open
subset $\CP^{3|3}:=\CPP^{3|3}\backslash \CPP^{1|3}$; its anti-self-dual
counterpart is $\CP^{3|3}_*\cong \CP^{3|3}$, where the parity assignment of the
appearing coordinates is simply inverted. Furthermore, we denote by $\CP^{2|3}$
the mini-supertwistor space obtained by dimensional reduction from $\CP^{3|3}$
and used in the description of the supersymmetric Bogomolny equations in three
dimensions.

For $\CN=4$ SYM theory, the appropriate twistor space $\CL^{5|6}$
is now obtained from the product $\CP^{3|3}\times \CP^{3|3}_*$ upon imposing a
quadric
condition reducing the bosonic dimensions by one\footnote{In fact, the field theory described by $\CL^{5|6}$ is $\CN=3$ SYM theory in four dimensions, which is equivalent to $\CN=4$ SYM theory on the level of equations of motion. In three dimensions, the same relation holds between $\CN=6$ and $\CN=8$ SYM theories.}. We perform an
analogous construction for $\CN=8$ SYM theory by starting from the
product $\CP^{2|3}\times\CP^{2|3}_*$ of two mini-supertwistor
spaces. The dimensional reduction turning the super self-duality equations in
four dimensions into the super Bogomolny equations in three dimensions
translates into a reduction of the quadric condition, which yields a constraint
only to be imposed on the diagonal
$\CPP^1_{\Delta}=\diag(\CPP^1\times \CPP^1_*)$ in the base of the
vector bundle $\CP^{2|3}\times \CP^{2|3}_*\rightarrow \CPP^1\times
\CPP^1_*$. Thus, the resulting space $\CL^{4|6}$ is not a vector
bundle but only a fibration and the sections of this fibration form a torsion
sheaf, as we will see. More explicitly, the
bosonic part of the fibres of $\CL^{4|6}$ over $\CPP^1\times
\CPP^1_*$ are isomorphic to $\FC^2$ at generic points, but over
the diagonal $\CPP^1_{\Delta}$, they are isomorphic to $\FC$.

As expected, we find a twistor correspondence between points in
$\FC^{3|12}$ and holomorphic sections of $\CL^{4|6}$ as well as
between points in $\CL^{4|6}$ and certain sub-supermanifolds in
$\FC^{3|12}$. After introducing a real structure on $\CL^{4|6}$,
one finds a nice interpretation of the spaces involved in the
twistor correspondence in terms of lines with marked points in
$\FR^3$, which resembles the appearance of flag manifolds in the
well-established twistor correspondences. Recalling that $\CL^{5|6}$ is a
Calabi-Yau supermanifold\footnote{the essential
prerequisite for being the target space of a topological B-model},
we are led to examine an analogous question for $\CL^{4|6}$. The Calabi-Yau
property essentially amounts to a vanishing of the first Chern
class of $T\CL^{5|6}$, which in turn encodes information about the
degeneracy locus of a certain set of sections of the vector bundle
$\CL^{5|6}\rightarrow \CPP^1\times \CPP^1_*$. We find that the
degeneracy loci of $\CL^{5|6}$ and $\CL^{4|6}$ are equivalent
(identical up to a principal divisor).

A Penrose-Ward transform for $\CN=8$ SYM theory can now be
conveniently established. To define the analogue of a holomorphic
vector bundle over the space $\CL^{4|6}$, we have to remember that
in the \v{C}ech description, a holomorphic vector bundle is
completely characterized by its transition functions, which in
turn form a group-valued \v{C}ech 1-cocycle. These objects are
still well-defined on $\CL^{4|6}$ and we will use such a
1-cocycle to define what we will call a pseudo-bundle over $\CL^{4|6}$. In performing the
transition between these pseudo-bundles and solutions to the
$\CN=8$ SYM equations, care must be taken when discussing these
bundles over the subset $\CL^{4|6}|_{\CPP^1_\Delta}$ of their
base. Eventually, however, one obtains a bijection between gauge
equivalence classes of solutions to the $\CN=8$ SYM equations and
equivalence classes of holomorphic pseudo-bundles over $\CL^{4|6}$, which turn into holomorphically trivial vector bundles upon restriction to
any holomorphic submanifold $\CPP^1\times \CPP^1_*\embd
\CL^{4|6}$.

Considering the reduction of $\CL^{5|6}\subset
\CP^{3|3}\times\CP^{3|3}_*$ to the bodies of the involved
spaces\footnote{i.e., putting the fermionic coordinates on all the
spaces to zero}, it is possible to find a twistor correspondence
for certain formal neighborhoods of $\CL^{5|0}\subset
\CP^{3|0}\times\CP^{3|0}_*$ on which a Penrose-Ward transform for
purely bosonic Yang-Mills theory in four dimensions can be built.
To improve our understanding of the mini-superambitwistor space it
is also helpful to discuss the analogous construction with
$\CL^{4|0}$. We find that a third order sub-thickening, i.e.\ a
thickening of the fibres which are only of dimension one, inside
of $\CP^{2|0}\times \CP^{2|0}_*$ must be considered to describe
solutions to the Yang-Mills-Higgs equations in three dimensions by
using pseudo-bundles over $\CL^{4|0}$.

To clarify the r{\^o}le of the space $\CL^{4|6}$ in detail, it would
be interesting to establish a dimensionally reduced version of the
construction considered by Movshev in \cite{Movshev:2004ub}. In
this paper, the author constructs a ``Chern-Simons triple''
consisting of a differential graded algebra $(A,\dd)$ and a
$\dd$-closed trace functional on a certain space $ST$ related to
the superambitwistor space. This Chern-Simons triple on $ST$ is
then conjectured to be equivalent to $\CN=4$ SYM theory in four
dimensions. The way the construction is performed suggests a quite
straightforward dimensional reduction to the case of the
mini-superambitwistor space. Besides delivering a Chern-Simons
triple for $\CN=8$ SYM theory in three dimensions, this
construction would possibly shed more light on the unusual
properties of the fibration $\CL^{4|6}$.

Following Witten's seminal paper \cite{Witten:2003nn}, there has
been growing interest in different supertwistor spaces suited as
target spaces for the topological B-model, see e.g.\
\cite{Popov:2004nk}-\cite{Chiou:2005jn}, \cite{Popov:2005uv}.
Although it is not clear what the topological B-model on
$\CL^{4|6}$ looks like exactly (we will present some speculations
in section 3.7), the mini-superambitwistor space might also prove to be
interesting from the topological string theory point of view.
In particular, the mini-superambitwistor space $\CL^{4|6}$ is
probably the mirror of the mini-supertwistor space $\CP^{2|4}$.
Maybe even the extension of infinite dimensional symmetry algebras
\cite{Wolf:2004hp} from the self-dual to the full case is easier
to study in three dimensions due to the greater similarity of
self-dual and full theory and the smaller number of conformal
generators.

Note that we are {\em not} describing the space of null geodesics in three
dimensions; this space has been constructed in \cite{LeBrun}.

The outline of this paper is as follows. In section 2, we review
the construction of the supertwistor spaces for SDYM theory and
SYM theory. Furthermore, we present the dimensional reduction
yielding the mini-supertwistor space used for capturing
solutions to the super Bogomolny equations. Section 3, the main
part, is then devoted to deriving the mini-superambitwistor space
in several ways and discussing in detail the associated twistor
correspondence and its geometry. Moreover, we comment on a
topological B-model on this space. In section 4, the Penrose-Ward
transform for three-dimensional $\CN=8$ SYM theory is presented.
First, we review both the transform for $\CN=4$ SYM theory in four
dimensions and aspects of $\CN=8$ SYM theory in three dimensions.
Then we introduce the pseudo-bundles over $\CL^{4|6}$, which take over the
r{\^o}le of vector bundles over the space $\CL^{4|6}$. Eventually, we
present the actual Penrose-Ward transform in detail. In the last
section, we discuss the third order sub-thickenings of $\CL^{4|0}$
in $\CP^{2|0}\times \CP^{2|0}_*$, which are used in the
Penrose-Ward transform for purely bosonic Yang-Mills-Higgs theory.

\section{Review of supertwistor spaces}

We shall briefly review some elementary facts on supertwistor
spaces and fix our conventions in this section. For a broader
discussion of supertwistor and superambitwistor spaces in
conventions close to the ones employed here, see
\cite{Popov:2004rb}. For more details on the mini-supertwistor
spaces, we refer to \cite{Chiou:2005jn} and \cite{Popov:2005uv}.

\subsection{Supertwistor spaces}

The supertwistor space of $\FC^{4|2\CN}$ is defined as the rank
$2|\CN$ holomorphic supervector bundle
\begin{equation}
\CP^{3|\CN}\ :=\ \FC^2\otimes \CO(1)\oplus \FC^\CN\otimes
\Pi\CO(1)
\end{equation}
over the Riemann sphere $\CPP^1$. Here, $\Pi$ is the parity
changing operator which inverts the parity of the fibre
coordinates. The base space of this bundle is covered by the two
patches $U_\pm$ on which we have the standard coordinates
$\lambda_\pm\in U_\pm\cong\FC$ with $\lambda_+=(\lambda_-)^{-1}$
on $U_+\cap U_-$. Over $U_\pm$, we introduce furthermore the
bosonic fibre coordinates $z^\alpha_\pm$ with $\alpha=1,2$ and the
fermionic fibre coordinates $\eta_i^\pm$ with $i=1,\ldots, \CN$.
On the intersection $U_+\cap U_-$, we thus have
\begin{equation}
z^\alpha_+\ =\ \frac{1}{z^3_-}\,\, z^\alpha_-\eand \eta_i^+\ =\
\frac{1}{z^3_-}\,\, \eta_i^-~~~\mbox{with}~~~z^3_\pm=\lambda_\pm~.
\end{equation}
The supermanifold $\CP^{3|\CN}$ as a whole is covered by the two
patches $\CU_\pm:=\CP^{3|\CN}|_{U_\pm}$ with local coordinates
$(z^1_\pm,z^2_\pm,z^3_\pm,\eta_1^\pm,\ldots,\eta_\CN^\pm)$.

Global holomorphic sections of the vector bundle
$\CP^{3|\CN}\rightarrow\CPP^1$ are given by polynomials of degree
one, which are parameterized by moduli
$(x^{\alpha\ald},\eta_i^\ald)\in\FC^{4|2\CN}$ via
\begin{equation}\label{incidence}
z^\alpha_\pm\ =\ x^{\alpha\ald} \lambda^\pm_\ald\eand \eta_i^\pm\
=\ \eta_i^\ald \lambda^\pm_\ald~,
\end{equation}
where we introduced the simplifying spinorial notation
\begin{equation}\label{spinorialnotation}
(\lambda_\ald^+)\ :=\  \left(\begin{array}{c} 1\\\lambda_+
\end{array}\right)\eand (\lambda_\ald^-)\ :=\ \left(\begin{array}{c}
\lambda_- \\ 1
\end{array}\right)~.
\end{equation}

The equations \eqref{incidence}, the so-called {\em incidence
relations}, define a twistor correspondence between the spaces
$\CP^{3|\CN}$ and $\FC^{4|2\CN}$, which can be depicted in the
double fibration
\begin{equation}\label{dblfibrationfourself}
\begin{aligned}
\begin{picture}(50,40)
\put(0.0,0.0){\makebox(0,0)[c]{$\CP^{3|\CN}$}}
%\put(32.0,0.1){\makebox(0,0)[c]{$\Leftrightarrow$}}
\put(64.0,0.0){\makebox(0,0)[c]{$\FC^{4|2\CN}$}}
\put(34.0,33.0){\makebox(0,0)[c]{$\CF^{5|2\CN}$}}
\put(7.0,18.0){\makebox(0,0)[c]{$\pi_2$}}
\put(55.0,18.0){\makebox(0,0)[c]{$\pi_1$}}
\put(25.0,25.0){\vector(-1,-1){18}}
\put(37.0,25.0){\vector(1,-1){18}}
\end{picture}
\end{aligned}
\end{equation}
Here, $\CF^{5|2\CN}\cong\FC^{4|2\CN}\times \CPP^1$ and the
projections are defined as
\begin{equation}
\pi_1(x^{\alpha\ald},\eta_i^\ald,\lambda_\ald^\pm)\ :=\
(x^{\alpha\ald},\eta_i^\ald)\eand
\pi_2(x^{\alpha\ald},\eta_i^\ald,\lambda_\ald^\pm)\ :=\
(x^{\alpha\ald}\lambda_\ald^\pm,\lambda_\pm,\eta_i^\ald\lambda^\pm_\ald)~.
\end{equation}
We can now read off the following correspondences:
\begin{equation}
\begin{aligned}
\left\{\,\mbox{projective lines $\CPP^1_{x,\eta}$ in
$\CP^{3|\CN}$}\right\}&\ \longleftrightarrow\
\left\{\,\mbox{points $(x,\eta)$ in $\FC^{4|2\CN}$}\right\}~,\\
\left\{\,\mbox{points $p$ in $\CP^{3|\CN}$}\right\}&\
\longleftrightarrow\ \left\{\,\mbox{null ($\beta$-)superplanes
$\FC_p^{2|2\CN}$ in $\FC^{4|2\CN}$}\right\}~.
\end{aligned}
\end{equation}
While the first correspondence is rather evident, the second one
deserves a brief remark. Suppose
$(\hat{x}^{\alpha\ald},\hat{\eta}_i^\ald)$ is a solution to the
incidence relations \eqref{incidence} for a fixed point $p\in
\CP^{3|\CN}$. Then the set of all solutions is given by
\begin{equation}\label{superplanes}
\{(x^{\alpha\ald},\eta_i^\ald)\}~~~\mbox{with}~~~ x^{\alpha\ald}\
=\ \hat{x}^{\alpha\ald}+\mu^\alpha\lambda^\ald_\pm\eand
\eta_i^\ald\ =\ \hat{\eta}_i^\ald+\eps_i \lambda_\pm^\ald~,
\end{equation}
where $\mu^\alpha$ is an arbitrary commuting 2-spinor and $\eps_i$
is an arbitrary vector with Gra{\ss}mann-odd entries. The coordinates
$\lambda^\ald_\pm$ are defined by \eqref{spinorialnotation} and
$\lambda^\ald_\pm:=\eps^{\ald\bed}\lambda_\bed^\pm$ with
$\eps^{\ed\zd}=-\eps^{\zd\ed}=1$. One can choose to work on any
patch containing $p$. The sets defined in \eqref{superplanes} are
then called {\em null} or {\em $\beta$-superplanes}.

The double fibration \eqref{dblfibrationfourself} is the
foundation of the Penrose-Ward transform between equivalence
classes of certain holomorphic vector bundles over $\CP^{3|\CN}$
and gauge equivalence classes of solutions to the $\CN$-extended
supersymmetric self-dual Yang-Mills equations on $\FC^4$, see
e.g.\ \cite{Popov:2004rb}.

The tangent spaces to the leaves of the projection $\pi_2$ are
spanned by the vector fields
\begin{equation}
V^\pm_\alpha\ :=\ \lambda^\ald_\pm\der{x^{\alpha\ald}}\eand
V^i_\pm\ :=\ \lambda^\ald_\pm\der{\eta^\ald_i}~.
\end{equation}
Note furthermore that $\CP^{3|4}$ is a Calabi-Yau supermanifold: The bosonic
fibres contribute each $+1$ to the first Chern class and the fermionic ones $-1$
(this is related to the fact that Berezin integration amounts to differentiating
with respect to a Gra{\ss}mann variable). Together with the contribution from
the tangent bundle of the base space, we have in total a trivial first Chern
class. This space is thus suited as the target space for a topological B-model
\cite{Witten:2003nn}.

\subsection{The superambitwistor space}

The idea leading naturally to a superambitwistor space is to
``glue together'' both the self-dual and anti-self-dual subsectors
of $\CN=3$ SYM theory to the full theory. For this, we obviously
need a twistor space $\CP^{3|3}$ with coordinates
$(z^\alpha_\pm,z^3_\pm,\eta_i^\pm)$ together with a ``dual''
copy\footnote{The word ``dual'' refers to the spinor indices and
{\em not} to the line bundles underlying $\CP^{3|3}$.}
$\CP^{3|3}_*$ with coordinates
$(u^\ald_\pm,u^3_\pm,\theta^i_\pm)$. The dual twistor
space is considered as a holomorphic supervector bundle over the
Riemann sphere $\CPP^1_*$ covered by the patches $U_\pm^*$ with
the standard local coordinates $\mu_\pm=u^{3}_\pm$. For
convenience, we again introduce the spinorial notation
$(\mu_\alpha^+)=(1,\mu_+)^T$ and $(\mu_\alpha^-)=(\mu_-,1)^T$. The
two patches covering $\CP^{3|3}_*$ will be denoted by
$\CU_\pm^*:=\CP^{3|3}_*|_{U_\pm^*}$ and the product space
$\CP^{3|3}\times \CP^{3|3}_*$ of the two supertwistor spaces is
thus covered by the four patches
\begin{equation}
\CU_{(1)}\ :=\ \CU_+\times\CU^*_+~,~~~ \CU_{(2)}\ :=\
\CU_-\times\CU^*_+~,~~~ \CU_{(3)}\ :=\ \CU_+\times\CU^*_-~,~~~
\CU_{(4)}\ :=\ \CU_-\times\CU^*_-~,
\end{equation}
on which we have the coordinates
$(z^\alpha_{(a)},z^3_{(a)},\eta_i^{(a)};u^\ald_{(a)},
u^{3}_{(a)},\theta^i_{(a)})$. This space is furthermore a
rank $4|6$ supervector bundle over the space $\CPP^1\times
\CPP^1_*$. The global sections of this bundle are parameterized by
elements of $\FC^{4|6}\times \FC^{4|6}_*$ in the following way:
\begin{equation}\label{ambisections}
z^\alpha_{(a)}\ =\
x^{\alpha\ald}\lambda_\ald^{(a)}~,~~~\eta^{(a)}_i\ =\
\eta_i^\ald\lambda_\ald^{(a)}~;~~~ u^\ald_{(a)}\ =\
x_*^{\alpha\ald}\mu_\alpha^{(a)}~,~~~ \theta^i_{(a)}\ =\
\theta^{\alpha i}\mu_\alpha^{(a)}~.
\end{equation}

The superambitwistor space is now the subspace $\CL^{5|6}\subset
\CP^{3|3}\times \CP^{3|3}_*$ obtained from the {\em quadric
condition} (the ``gluing condition'')
\begin{equation}\label{quadric2}
\kappa_{(a)}\ :=\
z^\alpha_{(a)}\mu_\alpha^{(a)}-u_{(a)}^\ald\lambda_\ald^{(a)}+
2\theta_{(a)}^i\eta_i^{(a)}\ =\ 0~.
\end{equation}
In the following, we will denote the restrictions of $\CU_{(a)}$
to $\CL^{5|6}$ by $\hat{\CU}_{(a)}$.

Because of the quadric condition \eqref{quadric2}, the bosonic
moduli are not independent on $\CL^{5|6}$, but one rather has the
relation
\begin{equation}\label{chiralcoords}
x^{\alpha\ald}\ =\ x^{\alpha\ald}_0-\theta^{\alpha
i}\eta_i^\ald~~~ \mbox{and}~~~x_*^{\alpha\ald}\ =\
x^{\alpha\ald}_0+\theta^{\alpha i}\eta^\ald_i~.
\end{equation}
The moduli $(x^{\alpha\ald})$ and $(x^{\alpha\ald}_*)$ are
therefore indeed anti-chiral and chiral coordinates on the
(complex) superspace $\FC^{4|12}$ and with this identification,
one can establish the following double fibration using equations
\eqref{ambisections}:
\begin{equation}\label{ambidblfibration2}
\begin{aligned}
\begin{picture}(80,40)
\put(0.0,0.0){\makebox(0,0)[c]{$\CL^{5|6}$}}
%\put(32.0,0.1){\makebox(0,0)[c]{$\Leftrightarrow$}}
\put(64.0,0.0){\makebox(0,0)[c]{$\FC^{4|12}$}}
\put(32.0,33.0){\makebox(0,0)[c]{$\CF^{6|12}$}}
\put(7.0,18.0){\makebox(0,0)[c]{$\pi_2$}}
\put(55.0,18.0){\makebox(0,0)[c]{$\pi_1$}}
\put(25.0,25.0){\vector(-1,-1){18}}
\put(37.0,25.0){\vector(1,-1){18}}
\end{picture}
\end{aligned}
\end{equation}
where $\CF^{6|12}\cong \FC^{4|12}\times\CPP^1\times\CPP^1_*$ and
$\pi_1$ is the trivial projection. Thus, one has the
correspondences
\begin{equation}
\begin{aligned}
\left\{\,\mbox{subspaces $(\CPP^1\times \CPP_*^1)_{x_0,\eta,\theta}$
in $\CL^{5|6}$} \right\}&\ \longleftrightarrow\
\left\{\,\mbox{points $(x_0,\eta,\theta)$ in $\FC^{4|12}$}\right\}~, \\
\left\{\,\mbox{points $p$ in $\CL^{5|6}$}\right\}&\
\longleftrightarrow\ \left\{\,\mbox{null superlines in
$\FC^{4|12}$}\right\}~.
\end{aligned}
\end{equation}
The above-mentioned null superlines are intersections of
$\beta$-superplanes and dual $\alpha$-super\-planes. Given a
solution $(\hat{x}^{\alpha\ald}_0,\hat{\eta}^\ald_i,\hat{\theta}^{\alpha i})$
to the incidence relations \eqref{ambisections} for a fixed point
$p$ in $\CL^{5|6}$, the set of points on such a null superline
takes the form
\begin{equation*}
\{(x^{\alpha\ald}_0,\eta^\ald_i,\theta^{\alpha i})\}~~\mbox{with}~~
x^{\alpha\ald}_0\ =\
\hat{x}^{\alpha\ald}_0+t\mu^\alpha_{(a)}\lambda^\ald_{(a)}~,~~
\eta^\ald_i\ =\ \hat{\eta}^\ald_i+\eps_i\lambda_{(a)}^\ald~,~~
\theta^{\alpha i}\ =\ \hat{\theta}^{\alpha
i}+\tilde{\eps}^i\mu^\alpha_{(a)}~.
\end{equation*}
Here, $t$ is an arbitrary complex number and $\eps_i$ and
$\tilde{\eps}^i$ are both 3-vectors with Gra{\ss}mann-odd components.
The coordinates $\lambda^\ald_{(a)}$ and $\mu^\alpha_{(a)}$ are
chosen from arbitrary patches on which they are both well-defined.
Note that these null superlines are in fact of dimension $1|6$.

The space $\CF^{6|12}$ is covered by four patches
$\tilde{\CU}_{(a)}:=\pi_2^{-1}(\hat{\CU}_{(a)})$ and the tangent
spaces to the $1|6$-dimensional leaves of the fibration
$\pi_2:~\CF^{6|12}\rightarrow\CL^{5|6}$ from
\eqref{ambidblfibration2} are spanned by the holomorphic vector
fields
\begin{align}\label{ambivec1}
W^{(a)}&:=\mu_{(a)}^\alpha\lambda_{(a)}^\ald\dpar_{\alpha\ald}~,~~~
D^i_{(a)}=\lambda_{(a)}^\ald D_\ald^i~~~\mbox{and}~~~
D^{(a)}_i=\mu_{(a)}^\alpha D_{\alpha i}~,
\end{align}
where $D_{\alpha i}$ and $D^i_\ald$ are the superderivatives
defined by
\begin{equation}
D_{\alpha i}\ :=\ \der{\theta^{\alpha
i}}+\eta_i^\ald\der{x^{\alpha\ald}_0}\eand D_\ald^i\ :=\
\der{\eta^\ald_i}+\theta^{\alpha i}\der{x^{\alpha\ald}_0}~.
\end{equation}

Just as the space $\CP^{3|4}$, the superambitwistor space
$\CL^{5|6}$ is a Calabi-Yau supermanifold. To prove this, note that we can count first Chern numbers with respect to the base $\CPP^1\times\CPP^1_*$ of $\CL^{5|6}$. In particular, we define the line bundle $\CO(m,n)$ to have first Chern numbers $m$ and $n$ with respect to the two $\CPP^1$s in the base. The (unconstrained) fermionic part of $\CL^{5|6}$ which is given by $\FC^3\otimes\Pi\CO(1,0)\oplus\FC^3\otimes\Pi\CO(0,1)$ contributes $(-3,-3)$ in this counting, which has to be cancelled by the body $\CL^5$ of $\CL^{5|6}$. Consider therefore the map
\begin{equation}
\kappa:(z^\alpha_{(a)},\lambda_\ald^{(a)},u^\ald_{(a)},\mu_\alpha^{(a)})
\mapsto \left(\left.\kappa_{(a)}\right|_{\eta=\theta=0},\lambda_\ald^{(a)},\mu_\alpha^{(a)}\right)~,
\end{equation}
where $\kappa_{(a)}$ has been defined in \eqref{quadric2}. This
map is a vector bundle morphism and gives rise to the short exact
sequence
\begin{equation}\label{ambiexactseq}
0\ \longrightarrow\ \CL^5\ \longrightarrow\
\FC^2\otimes\CO(1,0)\oplus\FC^2\otimes\CO(0,1)\
\stackrel{\kappa}{\longrightarrow}\ \CO(1,1)\ \longrightarrow\ 0~.
\end{equation}
The first Chern classes of the bundles in this sequence are elements of $H^2(\CPP^1\times\CPP^1,\RZ)\cong \RZ\times\RZ$, which we denote by $\alpha_1 h_1+\alpha_2 h_2$ with $\alpha_1,\alpha_2\in \RZ$. Then the short exact sequence \eqref{ambiexactseq} together with the Whitney product formula yields
\begin{equation}
(1+h_1)(1+h_1)(1+h_2)(1+h_2)\ =\ (1+\alpha_1 h_1+\alpha_2
h_2+\ldots)(1+ h_1+h_2)~,
\end{equation}
where $(\alpha_1,\alpha_2)$ label the first Chern class of $\CL^5$ considered as
a holomorphic vector bundle over $\CPP^1\times \CPP^1_*$. It follows that
$c_1=(1,1)$, and taking into account the contribution of the tangent space to
the base\footnote{Recall that $T^{1,0}\CPP^1 \cong \CO(2)$.} $\CPP^1\times
\CPP^1_*$, we conclude that the contribution of the tangent space to $\CL^5$ to the first Chern class of $\CL^{5|6}$ is cancelled by the contribution of the fermionic fibres.

Since $\CL^{5|6}$ is a Calabi-Yau supermanifold, this space can be
used as a target space for the topological B-model. However, it is
still unclear what the corresponding gauge theory action will look
like. The most obvious guess would be some holomorphic BF-type
theory \cite{holBFtheory} with B a ``Lagrange multiplier
(0,3)-form''.

\subsection{Reality conditions on the superambitwistor space}

On the supertwistor spaces $\CP^{3|\CN}$, one can define a real structure which leads to Kleinian signature on the body of the moduli space $\FR^{4|2\CN}$ of real holomorphic sections of the fibration $\pi_2$ in
\eqref{dblfibrationfourself}. Furthermore, if $\CN$ is even, one can can impose a symplectic Majorana condition which amounts to a second real structure which yields Euclidean signature. Above, we saw that the superambitwistor space $\CL^{5|6}$ originates from two copies of $\CP^{3|3}$ and therefore, we cannot straightforwardly impose the Euclidean reality condition. However, besides the real structure leading to Kleinian signature, one can additionally define a reality condition by relating spinors of opposite helicities to each other. In this way, we obtain a Minkowski metric on the body of $\FR^{4|12}$. In the following, we will focus on the latter.

Consider the anti-linear involution $\tau_M$ which acts on the coordinates of
$\CL^{5|6}$ according to
\begin{equation}\label{reality}
\tau_M(z^\alpha_\pm,\lambda_\ald^\pm,\eta_i^\pm;u^\ald,\mu_\alpha^\pm,
\theta^i_\pm)\
:=\
\left(\overline{u^\ald_\pm},\overline{\mu_\alpha^\pm},\overline{\theta^i_\pm};
\overline{z^\alpha_\pm},\overline{\lambda_\ald^\pm},\overline{\eta_i^\pm}
\right)~.
\end{equation}
Sections of the bundle $\CL^{5|6}\rightarrow \CPP^1\times
\CPP^1_*$ which are $\tau_M$-real are thus parameterized by the moduli
\begin{equation}
x^{\alpha\bed}\ =\ \overline{x^{\bed\alpha}}\eand \eta_i^\ald\ =\
\overline{\theta^{\alpha i}}~.
\end{equation}
We extract furthermore the contained real coordinates via the
identification
\begin{equation}\label{realcomponents}
\begin{aligned}
x^{1\dot{1}}&\ =\ x^0+x^1~,~~~&x^{1\dot{2}}&\ =\ x^2-\di x^3~,~\\
x^{2\dot{1}}&\ =\ x^2+\di x^3~,~~~&x^{2\dot{2}}&\ =\ x^0-x^1~,
\end{aligned}
\end{equation}
and obtain a metric of signature $(1,3)$ on $\FR^4$ from $\dd
s^2:=\det(\dd x^{\alpha\ald})$. Note that we can also make the
identification \eqref{realcomponents} in the complex case
$(x^\mu)\in\FC^4$, and then even on $\CP^{3|\CN}$. In the
subsequent discussion, we will always employ
\eqref{realcomponents} which is consistent, because we will not be interested in
the real version of $\CP^{3|\CN}$.

\subsection{The mini-supertwistor spaces}

To capture the situation obtained by a dimensional reduction
$\FC^{4|2\CN}\rightarrow \FC^{3|2\CN}$, one uses the so-called
mini-supertwistor spaces. Note that the vector field
\begin{equation}
\CT_3\ :=\ \der{x^3}\ =\
\di\left(\der{x^{2\ed}}-\der{x^{1\zd}}\right)
\end{equation}
considered on $\CF^{5|2\CN}$ from diagram \eqref{dblfibrationfourself} can be
split into a holomorphic and
an antiholomorphic part when restricted from $\CF^{5|2\CN}$ to
$\CP^{3|\CN}$:
\begin{equation}
\begin{aligned}
\CT_3|_{\CP^{3|\CN}}\ =\ \CT+\bar{\CT}~,~~~\CT_+\ =\
\di\left(\der{z^2_+}-z_+^3\der{z^1_+}\right)~,~~~\CT_-\ =\
\di\left(z_-^3\der{z^2_-}-\der{z^1_-}\right)~.
\end{aligned}
\end{equation}
Let $\CG$ be the abelian group generated by $\CT$. Then the orbit
space $\CP^{3|\CN}/\CG$ is given by the holomorphic supervector
bundle
\begin{equation}
\CP^{2|\CN}\ :=\ \CO(2)\oplus \FC^\CN\otimes \Pi\CO(1)
\end{equation}
over $\CPP^1$, and we call $\CP^{2|\CN}$ a {\em mini-supertwistor
space}. We denote the patches covering $\CP^{2|\CN}$ by
$\CV_\pm:=\CU_\pm\cap\CP^{2|\CN}$. The coordinates of the base and
the fermionic fibres of $\CP^{2|\CN}$ are the same as those of
$\CP^{3|\CN}$. For the bosonic fibres, we define
\begin{equation}\label{eq:3.10}
w_+^1\ :=\ z_+^1+z_+^3z_+^2~~~\mbox{on}~~~\CV_+\eand w_-^1\ :=\
z_-^2+z_-^3z_-^1~~~\mbox{on}~~~\CV_-~
\end{equation}
and introduce additionally $w^2_\pm:=z^3_\pm=\lambda_\pm$ for
convenience. On the intersection $\CV_+\cap\CV_-$, we thus have
the relation $w_+^1=(w^2_-)^{-2} w_-^1$. This implies that $w^1_\pm$ describes global sections of the line bundle $\CO(2)$. We parametrize these sections according to
\begin{equation}\label{incidencemini}
w^1_\pm\ =\
y^{\ald\bed}\lambda_{\ald}^\pm\lambda_{\bed}^\pm~~~\mbox{with}~~~
y^{\ald\bed}=y^{(\ald\bed)}\in\FC^3~,
\end{equation}
and the new moduli $y^{\ald\bed}$ are identified with the previous
ones $x^{\alpha\bed}$ by the equation
$y^{\ald\bed}=x^{(\ald\bed)}$. The incidence relation
\eqref{incidencemini} allows us to establish a double fibration
\begin{equation}\label{dblfibrationthreeself}
\begin{aligned}
\begin{picture}(50,40)
\put(0.0,0.0){\makebox(0,0)[c]{$\CP^{2|\CN}$}}
%\put(32.0,0.1){\makebox(0,0)[c]{$\Leftrightarrow$}}
\put(64.0,0.0){\makebox(0,0)[c]{$\FC^{3|2\CN}$}}
\put(34.0,33.0){\makebox(0,0)[c]{$\CK^{4|2\CN}$}}
\put(7.0,18.0){\makebox(0,0)[c]{$\nu_2$}}
\put(55.0,18.0){\makebox(0,0)[c]{$\nu_1$}}
\put(25.0,25.0){\vector(-1,-1){18}}
\put(37.0,25.0){\vector(1,-1){18}}
\end{picture}
\end{aligned}
\end{equation}
where $\CK^{4|2\CN}\cong \FC^{3|2\CN}\times \CPP^1$. We again
obtain a twistor correspondence
\begin{equation}
\begin{aligned}
\left\{\,\mbox{projective lines $\CPP^1_{x,\eta}$ in
$\CP^{2|\CN}$}\right\}\ \longleftrightarrow\ \left\{\,\mbox{points
$(y,\eta)$ in
$\FC^{3|2\CN}$}\right\}~,\hspace{0.2cm}\\
\left\{\,\mbox{points $p$ in $\CP^{2|\CN}$}\right\}\
\longleftrightarrow\ \left\{\,\mbox{$2|\CN$-dimensional
superplanes in $\FC^{3|2\CN}$}\right\}~.
\end{aligned}
\end{equation}
The $2|\CN$-dimensional superplanes in $\FC^{3|2\CN}$ are given by
the set
\begin{equation}
\{ (y^{\ald\bed},\eta^\ald_i) \}~~~\mbox{with}~~~ y^{\ald\bed}\ =\
\hat{y}^{\ald\bed}+\kappa^{(\ald}\lambda_\pm^{\bed)}~,~~~
\eta^\ald_i\ =\ \hat{\eta}^\ald_i+\eps_i\lambda_\pm^\ald~,
\end{equation}
where $\kappa^\ald$ and $\eps_i$ are an arbitrary complex 2-spinor
and a vector with Gra{\ss}mann-odd components, respectively. The point
$(\hat{y}^{\ald\bed},\hat{\eta}^\ald_i)\in\FC^{3|2\CN}$ is again
an initial solution to the incidence relations
\eqref{incidencemini} for a fixed point $p\in\CP^{2|\CN}$. Note
that although these superplanes arise from null superplanes in
four dimensions via dimensional reduction, they themselves are not
null.

The vector fields along the projection $\nu_2$ are now spanned by
\begin{equation}\label{vecfieldsmini}
V^\pm_\ald\ :=\ \lambda^\bed_\pm\dpar_{(\ald\bed)}\eand V^i_\pm\
:=\ \lambda^\ald_\pm\der{\eta^\ald_i}
\end{equation}
with
\begin{equation}\label{derivativesmini}
\dpar_{\ed\ed}:=\der{y^{\ed\ed}}~,~~~\dpar_{\zd\zd}:=\der{y^{\zd\zd}}\eand\dpar_
{(\ed\zd)}\
=\ \dpar_{(\zd\ed)}\ :=\ \frac{1}{2}\der{y^{\ed\zd}}~.
\end{equation}

The mini-supertwistor space $\CP^{2|4}$ is again a Calabi-Yau
supermanifold, and the gauge theory equivalent to the topological
B-model on this space is a holomorphic BF theory
\cite{Popov:2005uv}.

\section{The mini-superambitwistor space}

In this section, we define and examine the mini-superambitwistor
space $\CL^{4|6}$, which we will use to build a Penrose-Ward
transform involving solutions to $\CN=8$ SYM theory in three
dimensions. We will first give an abstract definition of
$\CL^{4|6}$ by a short exact sequence, and present more heuristic
ways of obtaining the mini-superambitwistor space later.

\subsection{Abstract definition of the mini-superambitwistor space}

The starting point is the product space $\CP^{2|3}\times
\CP^{2|3}_*$ of two copies of the $\CN=3$ mini-supertwistor space.
In analogy to the space $\CP^{3|3}\times \CP^{3|3}_*$, we have
coordinates
\begin{equation}\label{coordminisuperambi}
\left(w^1_{(a)},~w^2_{(a)}=\lambda_{(a)},~\eta_i^{(a)};~
v^1_{(a)},~v^2_{(a)}=\mu_{(a)},~\theta^i_{(a)}\right)
\end{equation}
on the patches $\CV_{(a)}$ which are Cartesian products of $\CV_\pm$ and
$\CV_\pm^*$:
\begin{equation}
\CV_{(1)}\ :=\ \CV_+\times\CV^*_+~,~~~ \CV_{(2)}\ :=\
\CV_-\times\CV^*_+~,~~~ \CV_{(3)}\ :=\ \CV_+\times\CV^*_-~,~~~
\CV_{(4)}\ :=\ \CV_-\times\CV^*_-~.
\end{equation}
For convenience, let us introduce the subspace $\CPP^1_{\Delta}$
of the base of the fibration $\CP^{2|3}\times
\CP^{2|3}_*\rightarrow \CPP^1\times \CPP^1_*$ as
\begin{equation}
\CPP^1_{\Delta}\ :=\ \diag(\CPP^1\times \CPP^1_*)\ =\
\{(\mu_\pm,\lambda_\pm)\in\CPP^1\times\CPP^1_*\,|\,\mu_\pm=\lambda_\pm\}~.
\end{equation}
Consider now the map $\xi:\CP^{2|3}\times \CP^{2|3}_*\rightarrow
\CO_{\CPP^1_{\Delta}}(2)$ which is defined by
\begin{equation}
\xi:
\left(w^1_{(a)},w^2_{(a)},\eta_i^{(a)};v^1_{(a)},v^2_{(a)},\theta^i_{(a)}
\right)\mapsto
\left\{\begin{array}{cl}
\left(w^1_\pm-v^1_\pm+2\theta^i_\pm\eta_i^\pm,\, w^2_\pm\right)
& \mbox{for } w^2_\pm=v^2_\pm \\
\left(0,\, w^2_{(a)}\right) & \mbox{else}
\end{array}\right.~,
\end{equation}
where $\CO_{\CPP^1_\Delta}(2)$ is the line bundle $\CO(2)$ over
$\CPP^1_\Delta$. In this definition, we used the fact that a point
for which $w^2_\pm=v^2_\pm $ is at least on one of the patches
$\CV_{(1)}$ and $\CV_{(4)}$. Note in particular, that the map
$\xi$ is a morphism of vector bundles. Therefore, we can define a
space $\CL^{4|6}$ via the short exact sequence
\begin{equation}\label{superminiambiexactseq}
0\ \longrightarrow\ \CL^{4|6}\ \longrightarrow\ \CP^{2|3}\times
\CP^{2|3}_*\ \stackrel{\xi}{\longrightarrow}\
\CO_{\CPP^1_{\Delta}}(2)\ \longrightarrow\ 0~,
\end{equation}
cf.\ \eqref{ambiexactseq}. We will call this space the {\em mini-superambitwistor space}. Analogously to above, we will denote the pull-back of the patches $\CV_{(a)}$
to $\CL^{4|6}$
by $\hat{\CV}_{(a)}$. Obviously, the space $\CL^{4|6}$ is a fibration, and we can switch to the corresponding short exact sequence of sheaves of local sections:
\begin{equation}\label{superminiambiexactseq2}
0\ \longrightarrow\ \CCL^{4|6}\ \longrightarrow\ \CCP^{2|3}\times
\CCP^{2|3}_*\ \stackrel{\xi}{\longrightarrow}\
\CCO_{\CPP^1_{\Delta}}(2)\ \longrightarrow\ 0~.
\end{equation}
Note the difference in notation: \eqref{superminiambiexactseq} is a sequence of vector bundles, while \eqref{superminiambiexactseq2} is a sequence of sheaves. To analyze the geometry of the space $\CL^{4|6}$ in more detail, we will restrict ourselves to the body of this space and put the fermionic coordinates to zero. Similarly to the case of the superambitwistor space, this is possible as the map $\xi$ does not affect the fermionic dimensions in the exact sequence \eqref{superminiambiexactseq}; this will become clearer in the discussion in section 3.2.

Inspired by the sequence defining the skyscraper sheaf\footnote{A sheaf with sections supported only at the point $p$.} $0\rightarrow \CCO(-1)\rightarrow \CCO\rightarrow \CCO_p\rightarrow 0$, we introduce the following short exact sequence:
\begin{equation}\label{modskyscraper}
0\ \longrightarrow\ \CCO(1,-1)\oplus\CCO(-1,1)\ \stackrel{\zeta}{\longrightarrow}\ \CCO(2,0)\oplus\CCO(0,2)\ \longrightarrow\
\CCO_{\CPP^1_{\Delta}}(2)\oplus\CCO_{\CPP^1_{\Delta}}(2)\ \longrightarrow\ 0~.
\end{equation}
Here, we defined $\zeta:(a,b)\mapsto(a\,\eps^{\ald\bed}\lambda_\ald\mu_\bed,b\,\eps^{\ald\bed}\lambda_\ald\mu_\bed)$, where $\lambda_\ald$ and $\mu_\ald$ are the usual homogeneous coordinates on the base space $\CPP^1\times\CPP^1_*$. The sheaf $\CCO_{\CPP^1_{\Delta}}(2)$ is a torsion sheaf (sometimes sloppily referred to as a skyscraper sheaf) with sections supported only over $\CPP^1_\Delta$. Finally, we trivially have the short exact sequence
\begin{equation}\label{trivseq}
0\ \longrightarrow\ \CCO_{\CPP^1_{\Delta}}(2)\ \stackrel{\alpha_1}{\longrightarrow}\ \CCO_{\CPP^1_{\Delta}}(2)\oplus\CCO_{\CPP^1_{\Delta}}(2)\ \stackrel{\alpha_2}{\longrightarrow}\
\CCO_{\CPP^1_{\Delta}}(2)\ \longrightarrow\ 0~,
\end{equation}
where $\alpha_1:(a)\mapsto(a,a)$ and $\alpha_2:(a,b)\mapsto(a-b)$.

Using the short exact sequences \eqref{superminiambiexactseq2}, \eqref{modskyscraper} and \eqref{trivseq} as well as the nine lemma, we can establish the following diagram:
\begin{equation*}
\begin{array}{ccccccccc}
× & × & 0 & × & 0 & × & 0 & × & × \\ 
× & × & \downarrow & × & \downarrow & × & \downarrow & × & × \\ 
0 & \rightarrow & \CCO(1,-1)\oplus\CCO(-1,1) & \stackrel{\zeta}{\rightarrow} & \CCL^{4} & \rightarrow & \CCO_{\CPP^1_{\Delta}}(2) & \rightarrow & 0 \\ 
× & × & \phantom{\id}\downarrow\id & × & \downarrow & × & \phantom{\alpha_1}\downarrow \alpha_1& × & × \\ 
0 & \rightarrow & \CCO(1,-1)\oplus\CCO(-1,1) & \stackrel{\zeta}{\rightarrow} & \CCO(2,0)\oplus\CCO(0,2) & \rightarrow & \CCO_{\CPP^1_{\Delta}}(2)\oplus\CCO_{\CPP^1_{\Delta}}(2) & \rightarrow & 0 \\ 
× & × & \downarrow & × & \downarrow & × & \phantom{\alpha_2}\downarrow \alpha_2 & × & × \\ 
0 & \rightarrow & 0 & \rightarrow & \CCO_{\CPP^1_{\Delta}}(2) & \stackrel{\id}{\rightarrow} & \CCO_{\CPP^1_{\Delta}}(2) & \rightarrow & 0 \\ 
× & × & \downarrow & × & \downarrow & × & \downarrow & × & × \\ 
× & × & 0 & × & 0 & × & 0 & × & ×
\end{array}
\end{equation*}

From the horizontal lines of this diagram and the five lemma, we conclude that $\CCL^{4}\oplus\CCO_{\CPP^1_{\Delta}}(2)=\CCO(2,0)\oplus\CCO(0,2)$. Thus, $\CCL^{4}$ is not a locally free sheaf\footnote{A more sophisticated argumentation would use the common properties of the torsion functor to establish that $\CCL^{4}$ is a torsion sheaf. Furthermore, one can write down a further diagram using the nine lemma which shows that $\CCL^{4}$ is a coherent sheaf.} but a torsion sheaf, whose stalks over $\CPP^1_{\Delta}$ are isomorphic to the stalks of
$\CCO_{\CPP^1_{\Delta}}(2)$, while the stalks over
$(\CPP^1\times\CPP^1_*)\backslash \CPP^1_{\Delta}$ are isomorphic to the stalks
of $\CCO(2,0)\oplus\CCO(0,2)$. Therefore, $\CL^{4}$ is not a vector bundle, but a fibration\footnote{The homotopy lifting property typically included in the definition of a fibration is readily derived from the definition of $\CL^{4}$.} with fibres $\FC^2$ over generic points and fibres $\FC$ over $\CPP^1_\Delta$. In particular, the total space of $\CL^{4}$ is {\em not} a manifold. 

The fact that the total space of the bundle $\CL^{4|6}$ is neither a supermanifold nor a supervector bundle over $\CPP^1\times \CPP^1_*$ seems at first slightly disturbing. However, we will show that once one is aware of this new aspect, it does not cause any deep difficulties as far as the twistor correspondence and the Penrose-Ward transform are concerned.

\subsection{The mini-superambitwistor space by dimensional
reduction}

In the following, we will motivate the abstract definition more concretely by considering explicitly the dimensional reduction of the space $\CL^{5|6}$; we will also fix our notation in terms of coordinates and moduli of sections. For this, we will first reduce the product space
$\CP^{3|3}\times \CP^{3|3}_*$ and then impose the appropriate
reduced quadric condition. In a first step, we want to
eliminate in both $\CP^{3|3}$ and $\CP^{3|3}_*$ the dependence on
the bosonic modulus $x^3$. Thus we should factorize by
\begin{equation}
\CT_{(a)}\ =\ \left\{\begin{array}{l}
\der{z^2_+}-z_+^3\der{z^1_+}~~~\mbox{on}~~~\CU_{(1)}\\
z_-^3\der{z^2_-}-\der{z^1_-}~~~\mbox{on}~~~\CU_{(2)}\\
\der{z^2_+}-z_+^3\der{z^1_+}~~~\mbox{on}~~~\CU_{(3)}\\
z_-^3\der{z^2_-}-\der{z^1_-}~~~\mbox{on}~~~\CU_{(4)}
\end{array}\right.\eand
\CT_{(a)}^*\ =\ \left\{\begin{array}{l}
\der{u^\zd_+}-u_+^3\der{u^\ed_+}~~~\mbox{on}~~~\CU_{(1)}\\
\der{u^\zd_+}-u_+^3\der{u^\ed_+}~~~\mbox{on}~~~\CU_{(2)}\\
u_-^3\der{u^\zd_-}-\der{u^\ed_-}~~~\mbox{on}~~~\CU_{(3)}\\
u_-^3\der{u^\zd_-}-\der{u^\ed_-}~~~\mbox{on}~~~\CU_{(4)}
\end{array}\right.~,
\end{equation}
which leads us to the orbit space
\begin{equation}
\CP^{2|3}\times \CP^{2|3}_*\ =\ (\CP^{3|3}/\CG)\times
(\CP^{3|3}_*/\CG^*)~,
\end{equation}
where $\CG$ and $\CG^*$ are the abelian groups generated by $\CT$
and $\CT^*$, respectively. Recall that the coordinates we use on
this space have been defined in \eqref{coordminisuperambi}. The
global sections of the bundle $\CP^{2|4}\times
\CP^{2|4}_*\rightarrow \CPP^1\times \CPP^1_*$ are captured by the
parametrization
\begin{equation}\label{miniambisections}
w^1_{(a)}\ =\
y^{\ald\bed}\lambda_\ald^{(a)}\lambda_\bed^{(a)}~,~~~ v^1_{(a)}\
=\ y^{\ald\bed}_*\mu_\ald^{(a)}\mu_\bed^{(a)}~,~~~\theta^i_{(a)}\
=\ \theta^{\ald i}\mu_\ald^{(a)}~,~~~ \eta^{(a)}_i\ =\
\eta_i^\ald\lambda_\ald^{(a)}~,
\end{equation}
where we relabel the indices of $\mu_\alpha^{(a)}\rightarrow
\mu_\ald^{(a)}$ and the moduli $y^{\alpha\beta}_*\rightarrow
y^{\ald\bed}_*$, $\theta^{i \alpha}\rightarrow\theta^{i \ald}$,
since there is no distinction between left- and right-handed
spinors on $\FR^3$ or its complexification $\FC^3$.

The next step is obviously to impose the quadric condition, gluing
together the self-dual and anti-self-dual parts. Note that when
acting with $\CT$ and $\CT^*$ on $\kappa_{(a)}$ as given in
\eqref{quadric2}, we obtain
\begin{equation}\label{actiononkappa}
\begin{aligned}
\CT_{(1)}\kappa_{(1)}&\ =\ \CT_{(1)}^*\kappa_{(1)}\ =\
(\mu_+-\lambda_+)~,& \CT_{(2)}\kappa_{(2)}&\ =\
\CT_{(2)}^*\kappa_{(2)}\ =\ (\lambda_-\mu_+-1)~,\\
\CT_{(3)}\kappa_{(3)}&\ =\ \CT_{(3)}^*\kappa_{(3)}\ =\
(1-\lambda_+\mu_-)~,&\CT_{(4)}\kappa_{(4)}&\ =\
\CT_{(4)}^*\kappa_{(4)}\ =\ (\lambda_--\mu_-)~.
\end{aligned}
\end{equation}
This implies that the orbits generated by $\CT$ and $\CT^*$ become
orthogonal to the orbits of $\der{\kappa}$ only at
$\mu_\pm=\lambda_\pm$. We can therefore safely impose the condition
\begin{equation}\label{quadricred}
\left.\left(w^1_{\pm}-v^1_{\pm}+
2\theta_{\pm}^i\eta_i^{\pm}\right)\right|_{\lambda_\pm=\mu_\pm} =\
0~,
\end{equation}
and the subset of $\CP^{2|3}\times \CP^{2|3}_*$ which satisfies
this condition is obviously identical to the mini-superambitwistor
space $\CL^{4|6}$ defined above.

The condition \eqref{quadricred} naturally fixes the
parametrization of global sections of the fibration $\CL^{4|6}$
by giving a relation between the moduli used in
\eqref{miniambisections}. This relation is completely analogous to
\eqref{chiralcoords} and reads
\begin{equation}\label{chiralredcoords}
y^{\ald\bed}\ =\ y^{\ald\bed}_0-\theta^{(\ald i}\eta_i^{\bed)}~~~
\mbox{and}~~~y_*^{\ald\bed}\ =\ y_0^{\ald\bed}+\theta^{(\ald
i}\eta^{\bed)}_i~.
\end{equation}
We clearly see that this parametrization arises from \eqref{chiralcoords} by
dimensional reduction from $\FC^4\rightarrow \FC^3$. Furthermore, even with this
identification, $w^1_\pm$ and $v^1_\pm$ are independent for $\lambda_\pm\neq
\mu_\pm$. Thus indeed, imposing the condition \eqref{quadricred} only at
$\lambda_\pm=\mu_\pm$ is the dimensionally reduced analogue of imposing the
condition \eqref{quadric2} on $\CP^{3|3}\times \CP^{3|3}_*$.

\subsection{Comments on further ways of constructing $\CL^{4|6}$}

Although the construction presented above seems most natural, one
can imagine other approaches of defining the space $\CL^{4|6}$.
Completely evident is a second way, which uses the description of
$\CL^{5|6}$ in terms of coordinates on $\CF^{6|12}$. Here, one
factorizes the correspondence space $\CF^{6|12}$ by the groups
generated by the vector field $\CT_3=\CT_3^*$ and obtains the
correspondence space $\CK^{5|12}\cong \FC^{3|12}\times
\CPP^1\times \CPP^1_*$ together with equation
\eqref{chiralredcoords}. A subsequent projection $\pi_2$ from the
dimensionally reduced correspondence space $\CK^{5|12}$ then
yields the mini-superambitwistor space $\CL^{4|6}$ as defined
above.

Furthermore, one can factorize $\CP^{3|3}\times \CP^{3|3}_*$ only
by $\CG$ to eliminate the dependence on one modulus. This will
lead to $\CP^{2|3}\times \CP^{3|3}_*$ and following the above
discussion of imposing the quadric condition on the appropriate
subspace, one arrives again at \eqref{quadricred} and the space
$\CL^{4|6}$. Here, the quadric condition already implies the
remaining factorization of $\CP^{2|3}\times\CP^{3|3}_*$ by
$\CG^*$.

Eventually, one could anticipate the identification of moduli in
\eqref{chiralredcoords} and therefore want to factorize by the
group generated by the combination $\CT+\CT^*$. Acting with this
sum on $\kappa_{(a)}$ will produce the sum of the results given in
\eqref{actiononkappa}, and the subsequent discussion of the
quadric condition follows the one presented above.

\subsection{Double fibration}

Knowing the parametrization of global sections of the
mini-superambitwistor space fibred over $\CPP^1\times \CPP^1_*$ as
defined in \eqref{chiralredcoords}, we can establish a double
fibration, similarly to all the other twistor spaces we
encountered so far. Even more instructive is the following
diagram, in which the dimensional reduction of the involved spaces
becomes evident:
\begin{equation}\label{ambdouble}
\begin{aligned}
\begin{picture}(100,85)
\put(0.0,0.0){\makebox(0,0)[c]{$\CL^{4|6}$}}
\put(0.0,52.0){\makebox(0,0)[c]{$\CL^{5|6}$}}
%\put(32.0,0.1){\makebox(0,0)[c]{$\Leftrightarrow$}}
\put(96.0,0.0){\makebox(0,0)[c]{$\FC^{3|12}$}}
\put(96.0,52.0){\makebox(0,0)[c]{$\FC^{4|12}$}}
\put(51.0,33.0){\makebox(0,0)[c]{$\CK^{5|12}$}}
\put(51.0,85.0){\makebox(0,0)[c]{$\CF^{6|12}$}}
\put(37.5,25.0){\vector(-3,-2){25}}
\put(55.5,25.0){\vector(3,-2){25}}
\put(37.5,77.0){\vector(-3,-2){25}}
\put(55.5,77.0){\vector(3,-2){25}}
\put(0.0,45.0){\vector(0,-1){37}}
\put(90.0,45.0){\vector(0,-1){37}}
\put(45.0,78.0){\vector(0,-1){37}}
\put(24.0,78.0){\makebox(0,0)[c]{$\pi_2$}}
\put(72.0,78.0){\makebox(0,0)[c]{$\pi_1$}}
\put(24.0,26.0){\makebox(0,0)[c]{$\nu_2$}}
\put(72.0,26.0){\makebox(0,0)[c]{$\nu_1$}}
\end{picture}
\end{aligned}
\end{equation}
The upper half is just the double fibration for the quadric
\eqref{ambidblfibration2}, while the lower half corresponds to the
dimensionally reduced case. The reduction of $\FC^{4|12}$ to
$\FC^{3|12}$ is obviously done by factoring out the
group generated by $\CT_3$. The same is true for the reduction of
$\CF^{6|12}\cong \FC^{4|12}\times\CPP^1\times\CPP^1_*$ to
$\CK^{5|12}\cong \FC^{3|12}\times\CPP^1\times\CPP^1_*$. The
reduction from $\CL^{5|6}$ to $\CL^{4|6}$ was given above and the
projection $\nu_2$ from $\CK^{5|12}$ onto $\CL^{4|6}$ is defined
by equations \eqref{miniambisections}. The four patches covering
$\CF^{6|12}$ will be denoted by
$\tilde{\CV}_{(a)}:=\nu_2^{-1}(\hat{\CV}_{(a)})$.

The double fibration defined by the projections $\nu_1$ and
$\nu_2$ yields the following twistor correspondences:
\begin{equation}\label{3.15}
\begin{aligned}
\left\{\,\mbox{subspaces $(\CPP^1\times \CPP^1)_{y_0,\eta,\theta}$
in $\CL^{4|6}$} \right\}&\ \longleftrightarrow\
\left\{\,\mbox{points $(y_0,\eta,\theta)$ in $\FC^{3|12}$}\right\}~, \\
\left\{\,\mbox{generic points $p$ in $\CL^{4|6}$}\right\}&\
\longleftrightarrow\ \left\{\,\mbox{superlines in
$\FC^{3|12}$}\right\}~,\\ \left\{\,\mbox{points $p$ in $\CL^{4|6}$
with $\lambda_\pm=\mu_\pm$}\right\}&\ \longleftrightarrow\
\left\{\,\mbox{superplanes in $\FC^{3|12}$}\right\}~.
\end{aligned}
\end{equation}
The superlines and the superplanes in $\FC^{3|12}$ are defined as the sets
\begin{equation*}
\begin{aligned}
\{(y^{\ald\bed},\eta_i^\ald,\theta^{\ald i})\}~~\mbox{with}~~
&y^{\ald\bed}\ =\
\hat{y}^{\ald\bed}+t\lambda_{(a)}^{(\ald}\mu_{(a)}^{\bed)}~,~&\eta^\ald_i\
=\ \hat{\eta}^\ald_i+\eps_i\lambda_{(a)}^\ald~,~&\theta^{\ald i}\
=\
\hat{\theta}^{\ald i}+\tilde{\eps}^i\mu_{(a)}^\ald~,\\
\{(y^{\ald\bed},\eta_i^\ald,\theta^{\ald i})\}~~\mbox{with}~~
&y^{\ald\bed}\ =\
\hat{y}^{\ald\bed}+\kappa^{(\ald}\lambda_{(a)}^{\bed)}~,~&\eta^\ald_i\
=\ \hat{\eta}^\ald_i+\eps_i\lambda_{(a)}^\ald~,~&\theta^{\ald i}\
=\ \hat{\theta}^{\ald i}+\tilde{\eps}^i\lambda^\ald_{(a)}~,
\end{aligned}
\end{equation*}
where $t$, $\kappa^\ald$, $\eps_i$ and $\tilde{\eps}^i$ are an
arbitrary complex number, a complex commuting 2-spinor and two
3-vectors with Gra{\ss}mann-odd components, respectively. Note that in
the last line, $\lambda^\ald_\pm=\mu_\pm^\ald$, and we could also
have written
\begin{equation*}
\{(y^{\ald\bed},\eta_i^\ald,\theta^{\ald i})\}~~~\mbox{with}~~~
y^{\ald\bed}\ =\
\hat{y}^{\ald\bed}+\kappa^{(\ald}\mu_{(a)}^{\bed)}~,~~~\eta^\ald_i\
=\ \hat{\eta}^\ald_i+\eps_i\mu_{(a)}^\ald~,~~~\theta^{\ald i}\ =\
\hat{\theta}^{\ald i}+\tilde{\eps}^i\mu_{(a)}^\ald~.
\end{equation*}

The vector fields spanning the tangent spaces to the leaves of the
fibration $\nu_2$ are for generic values of $\mu_\pm$ and
$\lambda_\pm$ given by
\begin{equation}\label{vecfieldsminiambi}
\begin{aligned}
W^{(a)}&\ :=\ \mu_{(a)}^\ald\lambda_{(a)}^\bed\dpar_{(\ald\bed)}~,\\
\tilde{D}^i_{(a)}&\ :=\ \lambda_{(a)}^\bed\tilde{D}^i_\bed\ :=\
\lambda_{(a)}^\bed\left(\der{\eta_i^\bed}+\theta^{\ald
i}\dpar_{(\ald\bed)}\right)~,\\
D^{(a)}_i&\ :=\ \mu_{(a)}^\ald D_{\ald i}\ :=\
\mu_{(a)}^\ald\left(\der{\theta^{\ald
i}}+\eta_i^\bed\dpar_{(\ald\bed)}\right)~,
\end{aligned}
\end{equation}
where the derivatives $\dpar_{(\ald\bed)}$ have been defined in
\eqref{derivativesmini}. At $\mu_\pm=\lambda_\pm$, however, the
fibres of the fibration $\CL^{4|6}$ over $\CPP^1\times \CPP^1_*$
loose one bosonic dimension. As the space $\CK^{5|12}$ is a
manifold, this means that this dimension has to become tangent to
the projection $\nu_2$. In fact, one finds that over
$\CPP^1_{\Delta}$, besides the vector fields given in
\eqref{vecfieldsminiambi}, also the vector fields
\begin{equation}
\tilde{W}^\pm_\bed\ =\ \mu^\ald_\pm\dpar_{(\ald\bed)}\ =\
\lambda^\ald_\pm\dpar_{(\ald\bed)}
\end{equation}
annihilate the coordinates on $\CL^{4|6}$. Therefore, the leaves
to the projection $\nu_2:\CK^{5|12}\rightarrow \CL^{4|6}$ are of
dimension $2|6$ for $\mu_\pm= \lambda_\pm$ and of dimension $1|6$
everywhere else.

\subsection{Real structure on $\CL^{4|6}$}

Quite evidently, a real structure on $\CL^{4|6}$ is inherited from
the one on $\CL^{5|6}$, and we obtain directly from
\eqref{reality} the action of $\tau_M$ on $\CP^{2|4}\times
\CP^{2|4}_*$, which is given by
\begin{equation}\label{realitymini}
\tau_M(w^1_\pm,\lambda_\ald^\pm,\eta_i^\pm;v^1_\pm,\mu_\ald^\pm,\theta^i_\pm)\
:=\
\left(\overline{v^1_\pm},\overline{\mu_\ald^\pm},\overline{\theta^i_\pm};
\overline{w^1_\pm},\overline{\lambda_\ald^\pm},\overline{\eta_i^\pm}\right)~.
\end{equation}
This action descends in an obvious manner to $\CL^{4|6}$, which
leads to a real structure on the moduli space $\FC^{3|12}$ via the
double fibration \eqref{ambdouble}. Thus, we have as the resulting
reality condition
\begin{equation}
y_0^{\ald\bed}\ =\ \overline{y_0^{\bed\ald}}\eand \eta_i^\ald\ =\
\overline{\theta^{\ald i}}~,
\end{equation}
and the identification of the bosonic moduli $y^{\ald\bed}$ with
the coordinates on $\FR^3$ reads as
\begin{equation}\label{realcomponentsmini}
\begin{aligned}
y_0^{\ed\ed}\ =\ x^0_0+x^1_0~,~~~y_0^{\ed\zd}\ =\ y_0^{\zd\ed}\ =\
x^2_0~,~~~y_0^{\zd\zd}\ =\ x^0_0-x^1_0~.
\end{aligned}
\end{equation}

The reality condition $\tau_M(\cdot)=\cdot$ is indeed fully
compatible with the condition \eqref{quadricred} which reduces
$\CP^{2|4}\times\CP^{2|4}_*$ to $\CL^{4|6}$. The base space
$\CPP^1\times \CPP^1_*$ of the fibration $\CL^{4|6}$ is reduced to
a single sphere $S^2$ with real coordinates
$\frac{1}{2}(\lambda_\pm+\mu_\pm)=\frac{1}{2}(\lambda_\pm+\bl_\pm)$
and
$\frac{1}{2\di}(\lambda_\pm-\mu_\pm)=\frac{1}{2\di}(\lambda_\pm-\bl_\pm)$,
while the diagonal $\CPP^1_\Delta$ is reduced to a circle
$S^1_\Delta$ parameterized by the real coordinates
$\frac{1}{2}(\lambda_\pm+\bl_\pm)$. The $\tau_M$-real sections of
$\CL^{4|6}$ have to satisfy $w^1_{\pm}=\tau_M(w^1_\pm)=\bar{v}^1_\pm$.
Thus, the fibres of the fibration
$\CL^{4|6}\rightarrow\CPP^1\times\CPP^1_*$, which are of complex
dimension $2|6$ over generic points in the base and complex
dimension $1|6$ over $\CPP^1_{\Delta}$, are reduced to fibres of
real dimension $2|6$ and $1|6$, respectively. In particular, note
that
$\theta^i_\pm\eta^\pm_i=\bar{\eta}^\pm_i\bar{\theta}^i_\pm=-\bar{\theta}
^i_\pm\bar{\eta}_i^\pm$
is purely imaginary and therefore the quadric condition
\eqref{quadricred} together with the real structure $\tau_M$
implies, that
$w^1_\pm=\bar{v}^1_\pm=\bar{w}^1_\pm+2\bar{\theta}^i_\pm\bar{\eta}^\pm_i$
for $\lambda_\pm=\mu_\pm=\bl_\pm$. Thus, the body $\z{w}^1_\pm$ of
$w^1_\pm$ is purely real and we have $w^1_\pm\ =\ \z{w}^1_\pm
-\theta^i_\pm\eta_i^\pm$ and $v^1_\pm\ =\ \z{w}^1_\pm
+\theta^i_\pm\eta_i^\pm$ on the diagonal $S^1_{\Delta}$.

\subsection{Interpretation of the involved real geometries}

For the best-known twistor correspondences, i.e.\ the
correspondence\footnote{more precisely: the compactified version thereof} \eqref{dblfibrationfourself}, its dual and the
correspondence \eqref{ambidblfibration2}, there is a nice
description in terms of flag manifolds, see e.g.\ \cite{Ward:vs}.
For the spaces involved in the twistor correspondences including
mini-twistor spaces, one has a similarly nice interpretation after
restricting to the real situation. For simplicity, we reduce our
considerations to the bodies\footnote{i.e.\ drop the fermionic
directions} of the involved geometries, as the extension to
corresponding supermanifolds is quite straightforward.

Let us first discuss the double fibration
\eqref{dblfibrationthreeself}, and assume that we have imposed a
suitable reality condition on the fibre coordinates, the details
of which are not important. We follow again the usual discussion
of the real case and leave the coordinates on the sphere complex.
As correspondence space on top of the double fibration, we have
thus the space $\FR^{3}\times S^2$, which we can understand as the
set of oriented lines\footnote{not only the ones through the
origin} in $\FR^3$ with one marked point. Clearly, the point of
such a line is given by an element of $\FR^3$, and the direction
of this line in $\FR^3$ is parameterized by a point on $S^2$. The
mini-twistor space $\CP^2\cong \CO(2)$ now is simply the space of
all lines in $\FR^3$ \cite{Hitchin:1982gh}. Similarly to the case
of flag manifolds, the projections $\nu_1$ and $\nu_2$ in
\eqref{dblfibrationthreeself} become therefore obvious. For
$\nu_1$, simply drop the line and keep the marked point. For
$\nu_2$, drop the marked point and keep the line. Equivalently, we
can understand $\nu_2$ as moving the marked point on the line to
its shortest possible distance from the origin. This leads to the
space $TS^2\cong \CO(2)$, where the $S^2$ parametrizes again the
direction of the line, which can subsequently be still moved
orthogonally to this direction, and this freedom is parameterized
by the tangent planes to $S^2$ which are isomorphic to $\FR^2$.

Now in the case of the fibration included in \eqref{ambdouble}, we
impose the reality condition \eqref{realitymini} on the fibre
coordinates of $\CL^4$. In the real case, the correspondence space
$\CK^5$ becomes the space $\FR^3\times S^2\times S^2$ and this is
the space of two oriented lines in $\FR^3$ intersecting in a
point. More precisely, this is the space of two oriented lines in
$\FR^3$ each with one marked point, for which the two marked
points coincide. The projections $\nu_1$ and $\nu_2$ in
\eqref{ambdouble} are then interpreted as follows. For $\nu_1$,
simply drop the two lines and keep the marked point. For $\nu_2$,
fix one line and move the marked point (the intersection point)
together with the second line to its shortest distance to the
origin. Thus, the space $\CL^4$ is the space of configurations in
$\FR^3$, in which a line has a common point with another line at
its shortest distance to the origin.

Let us summarize all the above findings in the following table:
\begin{center}
\begin{tabular}[h]{|c|l|}
\hline Space & Relation to $\FR^3$\\\hline
\hline $\FR^3$ & marked points in $\FR^3$
$\phantom{\underbrace{\overbrace{i}}}$\\
\hline $\FR^3\times S^2$ &
oriented lines with a marked point in $\FR^3$
$\phantom{\underbrace{\overbrace{i}}}$\\
\hline \raisebox{-.2cm}[0pt][0pt]{$\CP^2\cong\CO(2)$} & oriented
lines in $\FR^3$ (with a marked point at shortest
dis-$\phantom{\overbrace{i}}$\\[-0.06cm]
& tance to the origin.)$\phantom{\underbrace{i}}$\\
\hline $\FR^3\times S^2\times S^2$ & two oriented lines with a
common
marked point in $\FR^3$ $\phantom{\underbrace{\overbrace{i}}}$\\
\hline \raisebox{-.2cm}[0pt][0pt]{$\CL^4$} & two oriented lines
with a
common marked point at shortest$\phantom{\overbrace{i}}$\\[-0.06cm]
&distance from one of the lines to the origin in
$\FR^3$$\phantom{\underbrace{i}}$\\\hline
\end{tabular}
\end{center}

\subsection{Remarks concerning a topological B-model on $\CL^{4|6}$}

The space $\CL^{4|6}$ is not well-suited as a target space for a
topological B-model since it is not a (Calabi-Yau) manifold.
However, one clearly expects that it is possible to define an
analogous model since, if we assume that the conjecture in
\cite{Neitzke:2004pf} is correct, such a model should simply be
the mirror of the mini-twistor string theory considered in
\cite{Chiou:2005jn}. This model would furthermore yield some
holomorphic Chern-Simons type equations of motion. The latter
equations would then define holomorphic pseudo-bundles over $\CL^{4|6}$ by an analogue of a holomorphic structure. These bundles will be
introduced in section 4.3 and in our discussion, they substitute
the holomorphic vector bundles.

Interestingly, the space $\CL^{4|6}$ has a property which comes
close to vanishing of a first Chern class. Recall that for any
complex vector bundle, its Chern classes are Poincar{\'e} dual to the
degeneracy cycles of certain sets of sections (this is a
Gau{\ss}-Bonnet formula). More precisely, to calculate the first Chern
class of a rank $r$ vector bundle, one considers $r$ generic
sections and arranges them into an $r\times r$ matrix $L$. The
degeneracy loci on the base space are then given by the zero locus
of $\det(L)$. Clearly, this calculation can be translated directly
to $\CL^{4|6}$.

We will now show that $\CL^{4|6}$ and $\CL^{5|6}$ have equivalent
degeneracy loci, i.e.\ they are equal up to a principal divisor,
which, if we were speaking of ordinary vector bundles, would not
affect the first Chern class. Our discussion simplifies
considerably if we restrict our attention to the bodies of the two supertwistor
spaces and put all the fermionic coordinates to zero. Note that this will not
affect the result, as the quadric conditions defining $\CL^{5|6}$ and
$\CL^{4|6}$ do not affect the fermionic dimensions: the fermionic parts of the
fibrations $\CL^{5|6}$ and $\CL^{4|6}$ are identical, which is easily seen by
considering the global sections generating the total spaces of the fibrations.
Instead of the ambitwistor spaces, it is also easier to consider
the vector bundles $\CP^3\times \CP^3_*$ and $\CP^2\times\CP^2_*$
over $\CPP^1\times \CPP^1_*$, respectively, with the appropriately
restricted sets of sections. Furthermore, we will stick to our
inhomogeneous coordinates and perform the calculation only on the
patch $\CU_{(1)}$, but all this directly translates into
homogeneous, patch-independent coordinates. The matrices to be
considered are
\begin{equation*}
L_{\CL^5}\ =\ \left(\begin{array}{cccc} x_1^{1\ald}\lambda^+_\ald
& x_2^{1\ald}\lambda^+_\ald & x_3^{1\ald}\lambda^+_\ald &
x_4^{1\ald}\lambda^+_\ald \\
x_1^{2\ald}\lambda^+_\ald & x_2^{2\ald}\lambda^+_\ald &
x_3^{2\ald}\lambda^+_\ald &
x_4^{2\ald}\lambda^+_\ald \\
x_1^{\alpha 1}\mu^+_\alpha & x_2^{\alpha 1}\mu^+_\alpha  &
x_3^{\alpha 1}\mu^+_\alpha  &
x_4^{\alpha 1}\mu^+_\alpha \\
x_1^{\alpha 2}\mu^+_\alpha & x_2^{\alpha 2}\mu^+_\alpha  &
x_3^{\alpha 2}\mu^+_\alpha  & x_4^{\alpha 2}\mu^+_\alpha
\end{array}\right)\eand
L_{\CL^4}\ =\ \left(\begin{array}{cc}
y_1^{\ald\bed}\lambda^+_\ald\lambda^+_\bed &
y_2^{\ald\bed}\lambda^+_\ald\lambda^+_\bed \\
y_1^{\ald\bed}\mu^+_\ald\mu^+_\bed &
y_2^{\ald\bed}\mu^+_\ald\mu^+_\bed
\end{array}\right)~,
\end{equation*}
and one computes the degeneracy loci for generic moduli to be
given by the equations
\begin{equation}
(\lambda_+-\mu_+)^2\ =\
0\eand(\lambda_+-\mu_+)(\lambda_+-\varrho_+)\ =\ 0
\end{equation}
on the bases of $\CL^5$ and $\CL^4$, respectively. Here,
$\varrho_+$ is a rational function of $\mu_+$ and therefore it is
obvious that both degeneracy cycles are equivalent.

When dealing with degenerated twistor spaces, one usually retreats
to the correspondence space endowed with some additional symmetry
conditions \cite{Mason:rf}. It is conceivable that a similar
procedure will help to define the topological B-model in our case.
Also, defining a suitable blow-up of $\CL^{4|6}$ over
$\CPP^1_{\Delta}$ could be the starting point for finding an
appropriate action.

\section{The Penrose-Ward transform for the mini-superambitwistor space}

\subsection{Review of the Penrose-Ward transform on the
superambitwistor space}

Let $\CE$ be a topologically trivial holomorphic vector bundle of
rank $n$ over $\CL^{5|6}$ which becomes holomorphically trivial
when restricted to any subspace $(\CPP^1\times
\CPP^1)_{x_0,\eta,\theta}\embd\CL^{5|6}$. Due to the equivalence of
the \v{C}ech and the Dolbeault descriptions of holomorphic vector
bundles, we can describe $\CE$ either by holomorphic transition
functions $\{f_{ab}\}$ or by a holomorphic structure
$\dparb_{\hat{\CA}}=\dparb+\hat{\CA}$: Starting from a transition
function $f_{ab}$, there is a splitting
\begin{equation}
f_{ab}\ =\ \hat{\psi}_a^{-1}\hat{\psi}_b~,
\end{equation}
where the $\hat{\psi}_a$ are smooth $\sGL(n,\FC)$-valued
functions\footnote{In fact, the collection $\{\hat{\psi}_a\}$
forms a \v{C}ech 0-cochain.} on $\CU_{(a)}$, since the bundle
$\CE$ is topologically trivial. This splitting allows us to switch
to the holomorphic structure $\dparb+\hat{\CA}$ with
$\hat{\CA}=\hat{\psi}\dparb\hat{\psi}^{-1}$, which describes a
trivial vector bundle $\hat{\CE}\cong \CE$. Note that the
additional condition of holomorphic triviality of $\CE$ on
subspaces $(\CPP^1\times \CPP^1)_{x_0,\eta,\theta}$ will restrict
the explicit form of $\hat{\CA}$.

Back at the bundle $\CE$, consider its pull-back $\pi_2^*\CE$ with
transition functions $\{\pi_2^*f_{ab}\}$, which are constant along
the fibres of $\pi_2:\CF^{6|12}\rightarrow\CL^{5|6}$:
\begin{equation}\label{ambWcond1}
W^{(a)}\pi_2^*f_{ab}\ =\ D^i_{(a)}\pi_2^*f_{ab}\ =\
D_i^{(a)}\pi_2^*f_{ab}\ =\ 0~,
\end{equation}
The additional assumption of holomorphic triviality upon reduction
onto a subspace allows for a splitting
\begin{equation}\label{splittinghol}
\pi_2^*f_{ab}\ =\ \psi_a^{-1}\psi_b
\end{equation}
into $\sGL(n,\FC)$-valued functions $\{\psi_a\}$ which are
holomorphic on $\tilde{\CU}_{(a)}$: Evidently, there is such a
splitting holomorphic in the coordinates $\lambda_{(a)}$ and
$\mu_{(a)}$ on $(\CPP^1\times \CPP^1)_{x_0,\eta,\theta}$, since
$\CE$ becomes holomorphically trivial when restricted to these
spaces. Furthermore, these subspaces are holomorphically
parameterized by the moduli $(x^{\alpha\ald}_0,\eta^\ald_i,\theta^{\alpha i})$,
and thus the splitting \eqref{splittinghol} is holomorphic in all the
coordinates of $\CF^{6|12}$. Due to \eqref{ambWcond1}, we have on the
intersections $\tilde{\CU}_{(a)}\cap\tilde{\CU}_{(b)}$
\begin{subequations}
\begin{align}\label{ambpbgauge5}
\psi_a D^i_{(a)}\psi^{-1}_a\ =\ \psi_b D^i_{(a)}\psi^{-1}_b&\ =:\
\lambda^\ald_{(a)}\CA_\ald^i~,\\\label{ambpbgauge6} \psi_a
D_i^{(a)}\psi^{-1}_a\ =\ \psi_b D_i^{(a)}\psi^{-1}_b&\ =:\
\mu^\alpha_{(a)}\CA_{\alpha i}~,\\\label{ambpbgauge7} \psi_a
W^{(a)}\psi^{-1}_a\ =\ \psi_b W^{(a)}\psi^{-1}_b&\ =:\
\mu^\alpha_{(a)} \lambda^\ald_{(a)}\CA_{\alpha\ald}~,
\end{align}
\end{subequations}
where $\CA_\ald^i$, $\CA_{\alpha i}$ and $\CA_{\alpha\ald}$ are
independent of $\mu_{(a)}$ and $\lambda_{(a)}$. The introduced
components of the supergauge potential $\CA$ fit into the linear
system
\begin{subequations}
\begin{align}\label{amblinsys21}
\mu_{(a)}^\alpha\lambda_{(a)}^\ald(\dpar_{\alpha\ald}+
\CA_{\alpha\ald})\psi_a&\ =\ 0~,\\
\lambda_{(a)}^\ald(D^i_{\ald}+\CA^i_{\ald})\psi_a&\ =\
0~,\\\label{amblinsys24} \mu_{(a)}^\alpha(D_{\alpha i}+\CA_{\alpha
i})\psi_a&\ =\ 0~,
\end{align}
\end{subequations}
whose compatibility conditions are
\begin{equation}\label{ambcompcond1}
\begin{aligned}
\{\nabla_\ald^i,\nabla_\bed^j\}+\{\nabla_\bed^i,\nabla_\ald^j\}=0~,~~~
\{\nabla_{\alpha i},\nabla_{\beta j}\}+\{\nabla_{\beta i},
\nabla_{\alpha j}\}=0~,\\ \{\nabla_{\alpha
i},\nabla^j_{\ald}\}-2\delta_i^j\nabla_{\alpha\ald}=0~.\hspace{3cm}
\end{aligned}
\end{equation}
Here, we used the obvious shorthand notations
$\nabla_\ald^i:=D_\ald^i+\CA_\ald^i$, $\nabla_{\alpha
i}:=D_{\alpha i}+\CA_{\alpha i}$, and
$\nabla_{\alpha\ald}=\dpar_{\alpha\ald}+\CA_{\alpha\ald}$.
Equations \eqref{ambcompcond1} are well known to be equivalent to
the equations of motion of $\CN=3$ SYM theory on\footnote{Note
that most of our considerations concern the complexified case.}
$\FC^4$ \cite{Witten:1978xx}, and therefore also to $\CN=4$ SYM theory on
$\FC^4$.

We thus showed that there is a correspondence between certain
holomorphic structures on $\CL^{5|6}$, holomorphic vector bundles
over $\CL^{5|6}$ which become holomorphically trivial when
restricted to certain subspaces and solutions to the $\CN=4$ SYM
equations on $\FC^4$. The redundancy in each set of objects is
modded out by considering gauge equivalence classes and
holomorphic equivalence classes of vector bundles, which renders
the above correspondences one-to-one.

\subsection{$\CN=8$ SYM theory in three dimensions}

This theory is obtained by dimensionally reducing $\CN=1$ SYM
theory in ten dimensions to three dimensions, or, equivalently, by
dimensionally reducing four-dimensional $\CN=4$ SYM theory to
three dimensions. As a result, the 16 real supercharges are
re-arranged in the latter case from four spinors transforming as a
$\mathbf{2}_\FC$ of $\sSpin(3,1)\cong\sSL(2,\FC)$ into eight
spinors transforming as a $\mathbf{2}_\FR$ of $\sSpin(2,1)\cong
\sSL(2,\FR)$.

The automorphism group of the supersymmetry algebra is
$\sSpin(8)$, and the little group of the remaining Lorentz group
$\sSO(2,1)$ is trivial. As massless particle content, we therefore
expect bosons transforming in the $\mathbf{8}_v$ and fermions
transforming in the $\mathbf{8}_c$ of $\sSpin(8)$. One of the
bosons will, however, appear as a dual gauge potential on $\FR^3$
after dimensional reduction, and therefore only a $\sSpin(7)$
R-symmetry group is manifest in the action and the equations of
motion. In the mini-superambitwistor formulation, the manifest
subgroup of the R-symmetry group is only $\sSU(3)\times
\sU(1)\times \sSU(3)\times \sU(1)$. Altogether, we have a gauge
potential $A_\mu$ with $\mu=1,\ldots,3$, seven scalars $\phi^i$
with $i=1,\ldots,7$ and eight spinors $\chi^j_\ald$ with
$j=1,\ldots,8$.

Moreover, recall that in four dimensions, $\CN=3$ and $\CN=4$
super Yang-Mills theories are equivalent on the level of field
content and corresponding equations of motion. The only
difference is found in the manifest R-symmetry groups
which are $\sSU(3)\times \sU(1)$ and $\sSU(4)$, respectively. This
equivalence obviously carries over to the three-dimensional
situation: $\CN=6$ and $\CN=8$ super Yang-Mills theories are
equivalent regarding their field content and the equations of
motion. Therefore, it is sufficient to construct a twistor
correspondence for $\CN=6$ SYM theory to describe solutions to the
$\CN=8$ SYM equations.

\subsection{Pseudo-bundles over $\CL^{4|6}$}

Because the mini-superambitwistor space is only a fibration and
not a manifold, there is no notion of holomorphic vector bundles
over $\CL^{4|6}$. However, our space is close enough to a manifold
to translate all the necessary terms in a simple manner.

Let us fix the covering $\frU$ of the total space of the fibration
$\CL^{4|6}$ to be given by the patches $\CV_{(a)}$ introduced
above. Furthermore, define $\frS$ to be the sheaf of smooth
$\sGL(n,\FC)$-valued functions on $\CL^{4|6}$ and $\frH$ to be its
subsheaf consisting of holomorphic $\sGL(n,\FC)$-valued functions
on $\CL^{4|6}$, i.e.\ smooth and holomorphic functions which
depend only on the coordinates given in \eqref{miniambisections}
and $\lambda_{(a)},\mu_{(a)}$.

We define a {\em complex pseudo-bundle over $\CL^{4|6}$} of rank $n$ by a
\v{C}ech 1-cocycle $\{f_{ab}\}\in \check{Z}^1(\frU,\frS)$ on
$\CL^{4|6}$ in full analogy with transition functions defining
ordinary vector bundles. If the 1-cocycle is an element of
$\check{Z}^1(\frU,\frH)$, we speak of a {\em holomorphic
pseudo-bundle over $\CL^{4|6}$}. Two pseudo-bundles given by \v{C}ech
1-cocycles $\{f_{ab}\}$ and $\{f'_{ab}\}$ are called {\em
topologically equivalent} ({\em holomorphically equivalent}), if
there is a \v{C}ech 0-cochain $\{\psi_{a}\}\in
\check{C}^0(\frU,\frS)$ (a \v{C}ech 0-cochain $\{\psi_{a}\}\in
\check{C}^0(\frU,\frH)$) such that
$f_{ab}=\psi_a^{-1}f'_{ab}\psi_b$. A pseudo-bundle is called
{\em trivial} ({\em holomorphically trivial}), if it is
topologically equivalent (holomorphically equivalent) to the
trivial pseudo-bundle given by $\{f_{ab}\}=\{\unit_{ab}\}$.

In the corresponding discussion of \v{C}ech cohomology on ordinary
manifolds, one can achieve independence of the covering if the
patches of the covering are all Stein manifolds. An analogous
argument should also be applicable here, but for our purposes, it
is enough to restrict to the covering $\frU$.

Besides the \v{C}ech description, it is also possible to introduce
an equivalent Dolbeault description, which will, however, demand
an extended notion of Dolbeault cohomology classes.

\subsection{The Penrose-Ward transform using the
mini-superambitwistor space}

With the double fibration contained in \eqref{ambdouble}, it is
not hard to establish the corresponding Penrose-Ward transform,
which is essentially a dimensional reduction of the
four-dimensional case presented in section 4.1.

On $\CL^{4|6}$, we start from a trivial rank $n$ holomorphic
pseudo-bundle over $\CL^{4|6}$ defined by a 1-cocycle $\{f_{ab}\}$ which
becomes a holomorphically trivial vector bundle upon restriction
to any subspace $(\CPP^1\times \CPP^1)_{y_0,\eta,\theta}\embd
\CL^{4|6}$. The pull-back of the pseudo-bundle over $\CL^{4|6}$ along $\nu_2$
is the vector bundle $\tilde{\CE}$ with transition functions
$\{\nu_2^* f_{ab}\}$ satisfying by definition
\begin{equation}\label{ambWcond13d}
W^{(a)}\nu_2^*f_{ab}\ =\ \tilde{D}^i_{(a)}\nu_2^*f_{ab}\ =\
D_i^{(a)}\nu_2^*f_{ab}\ =\ 0~,
\end{equation}
at generic points of $\CL^{4|6}$ and for $\lambda_\pm=\mu_\pm$, we
have
\begin{equation}\label{ambWcond13Deltad}
\tilde{W}^{(a)}_\ald\nu_2^*f_{ab}\ =\
\tilde{D}^i_{(a)}\nu_2^*f_{ab}\ =\ D_i^{(a)}\nu_2^*f_{ab}\ =\ 0~.
\end{equation}
Restricting the bundle $\tilde{\CE}$ to a subspace $(\CPP^1\times
\CPP^1)_{y_0,\eta,\theta}\embd \CL^{4|6}\subset \CF^{5|12}$ yields a
splitting of the transition function $\nu_2^*f_{ab}$
\begin{equation}
\nu_2^*f_{ab}\ =\ \psi_a^{-1}\psi_b~,
\end{equation}
where $\{\psi_a\}$ are again $\sGL(n,\FC)$-valued functions on
$\tilde{\CV}_{(a)}$ which are holomorphic. From this splitting
together with \eqref{ambWcond13d}, one obtains that
\begin{equation}\label{ambpbgauge}
\begin{aligned}
\psi_a \tilde{D}^i_{(a)}\psi^{-1}_a\ =\ \psi_b
\tilde{D}^i_{(a)}\psi^{-1}_b&\ =:\
\lambda^\ald_{(a)}\tilde{\CA}_\ald^i~,\\
\psi_a D_i^{(a)}\psi^{-1}_a\ =\ \psi_b D_i^{(a)}\psi^{-1}_b&\ =:\
\mu^\ald_{(a)}\CA_{\ald i}~,\\\psi_a W^{(a)}\psi^{-1}_a\ =\ \psi_b
W^{(a)}\psi^{-1}_b&\ =:\ \mu^\ald_{(a)}
\lambda^\bed_{(a)}\CB_{\ald\bed}~~\mbox{for}~~\lambda\neq\mu~,\\\psi_a
W^{(a)}_\ald\psi^{-1}_a\ =\ \psi_b
W^{(a)}_\ald\psi^{-1}_b&\ =:\ 
\lambda^\bed_{(a)}\tilde{\CB}_{\ald\bed}~~\mbox{for}~~\lambda=\mu~.
\end{aligned}
\end{equation}
These equations are due to a generalized Liouville theorem, and continuity
yields that $\tilde{\CB}_{\ald\bed}=\CB_{\ald\bed}$. Furthermore, one
immediately notes that a transition function
$\nu_2^* f_{ab}$, which satisfies \eqref{ambWcond13d} is of the
form
\begin{equation}
f_{ab}\ =\
f_{ab}(y^{\ald\bed}\lambda^{(a)}_{\ald}\lambda^{(a)}_{\bed},
y^{\ald\bed}\mu^{(a)}_{\ald}\mu^{(a)}_{\bed},\lambda^{(a)}_{\ald},\mu^{(a)}_{
\ald})~,
\end{equation}
and thus condition \eqref{ambWcond13Deltad} is obviously
fulfilled at points with $\lambda_\pm=\mu_\pm$. Altogether, since we neither
loose any information on the gauge potential nor do we loose any constraints on
it, we can restrict our discussion to generic points with $\lambda\neq \mu$,
which simplifies the presentation.

The superfield $\CB_{\ald\bed}$ decomposes into a
gauge potential and a Higgs field $\Phi$:
\begin{equation}
\CB_{\ald\bed}\ :=\
\CA_{(\ald\bed)}+\tfrac{\di}{2}\eps_{\ald\bed}\Phi~.
\end{equation}
The zeroth order component in the superfield expansion of $\Phi$
will be the seventh real scalar joining the six scalars of $\CN=4$
SYM in four dimensions, which are the zeroth component of the
superfield $\Phi_{ij}$ defined in
\begin{equation}
\{D_{\ald i}+\CA_{\ald i},D_{\bed j}+\CA_{\bed j}\}\ =:\
-2\eps_{\ald\bed}\Phi_{ij}~.
\end{equation}
Thus, as mentioned above, the $\sSpin(7)$ R-symmetry group of
$\CN=8$ SYM theory in three dimensions will not be manifest in
this description.

The equations \eqref{ambpbgauge} are equivalent to the linear
system
\begin{equation}\label{amblinsys}
\begin{aligned}
\mu_{(a)}^\ald\lambda_{(a)}^\bed(\dpar_{(\ald\bed)}+
\CB_{\ald\bed})\psi_a&\ =\ 0~,\\
\lambda_{(a)}^\ald(\tilde{D}^i_{\ald}+\tilde{\CA}^i_{\ald})\psi_a&\
=\ 0~,\\ \mu_{(a)}^\ald(D_{\ald i}+\CA_{\ald i})\psi_a&\ =\ 0~.
\end{aligned}
\end{equation}
To discuss the corresponding compatibility conditions, we
introduce the following differential operators:
\begin{equation}
\begin{aligned}
\tilde{\nabla}_\ald^i\ :=\
\tilde{D}_\ald^i+\tilde{\CA}_\ald^i~,~~~
\nabla_{\ald i}\ :=\ D_{\ald i}+\CA_{\ald i}~,\\
\nabla_{\ald\bed}\ :=\
\dpar_{(\ald\bed)}+\CB_{\ald\bed}~.\hspace{1.7cm}
\end{aligned}
\end{equation}
We thus arrive at
\begin{equation}\label{ambcompcond2}
\begin{aligned}
&\{\tilde{\nabla}_\ald^i,\tilde{\nabla}_\bed^j\}+\{\tilde{\nabla}_\bed^i,\tilde{
\nabla}_\ald^j\}\
=\ 0~,~~~ \{\nabla_{\ald i},\nabla_{\bed j}\}+\{\nabla_{\bed i},
\nabla_{\ald j}\}\ =\ 0~,\\ &\hspace{3cm}\{\nabla_{\ald
i},\tilde{\nabla}^j_{\bed}\}-2\delta_i^j\nabla_{\ald\bed}\ =\ 0~,
\end{aligned}
\end{equation}
and one clearly sees that equations \eqref{ambcompcond2} are
indeed equations \eqref{ambcompcond1} after a dimensional
reduction $\FC^4\rightarrow \FC^3$ and defining $\Phi:=A_4$. As it
is well known, the supersymmetry (and the R-symmetry) of $\CN=4$
SYM theory are enlarged by this dimensional reduction and we
therefore obtained indeed $\CN=8$ SYM theory on $\FC^3$.

To sum up, we obtained a correspondence between holomorphic
pseudo-bundles over $\CL^{4|6}$ which become holomorphically trivial vector
bundles upon reduction to any subspace $(\CPP^1\times
\CPP^1)_{y_0,\eta,\theta}\embd\CL^{4|6}$ and solutions to the
three-dimensional $\CN=8$ SYM equations. As this correspondence
arises by a dimensional reduction of a correspondence which is
one-to-one, it is rather evident, that also in this case, we have
a bijection between both the holomorphic pseudo-bundles over $\CL^{4|6}$ and the solutions after factoring out holomorphic equivalence and gauge equivalence, respectively.

\section{Purely bosonic Yang-Mills-Higgs theory from third order
subneighborhoods}

In this section, we want to turn to the purely bosonic
situation\footnote{In other words, all the superspaces used up to
now loose their Gra{\ss}mann-odd dimensions.} and describe solutions
to the three-dimensional Yang-Mills-Higgs\footnote{When speaking
about Yang-Mills-Higgs theory, we mean a theory without quartic
interaction term.} equations using a mini-ambitwistor space. That
is, we will consider the dimensional reduction of the purely
bosonic case discussed in \cite{Witten:1978xx} and
\cite{Isenberg:kk} from $d=4$ to $d=3$. In these papers, it has
been shown that solutions to the Yang-Mills field equations are
equivalent to holomorphic vector bundles over a third order
thickening of the ambitwistor space $\CL^{5}$ in $\CP^3\times
\CP^3_*$.

\subsection{Thickenings of complex manifolds}

Given a complex manifold $Y$ of dimension $d$, a thickening
\cite{Eastwood:1986} of a submanifold $X\subset Y$ with
codimension 1 is an infinitesimal neighborhood of $X$ in $Y$
described by an additional Gra{\ss}mann-even but nilpotent coordinate.
More precisely, the $m$-th order thickening of $X$ is denoted by
$X_{(m)}$ and defined as the manifold $X$ together with the
extended structure sheaf
\begin{equation}
\CO_{(m)}=\CO_Y/\CI^{m+1}~,
\end{equation}
where $\CO_Y$ is the structure sheaf of $Y$ and $\CI$ the ideal of
functions on $Y$ which vanish on $X$. We can choose local
coordinates $(x^1,\ldots,x^{d-1},y)$ on $Y$ such that $X$ is given
by $y=0$. The $m$-th order thickening $X_{(m)}$ given by the
scheme $(X,\CO_{(m)})$ is then described by the coordinates
$(x^1,\ldots,x^{d-1},y)$ together with the relation $y^{m+1}\sim
0$. For more details on infinitesimal neighborhoods, see e.g.\
\cite{Saemann:2004tt} and references therein.

Note that it is easily possible to map $\CL^{5|6}$ to a third
order thickening of $\CL^5\subset\CP^3\times\CP^3_*$ by
identifying the nilpotent even coordinate $y$ with
$2\theta^i\eta_i$, cf.\ \cite{Eastwood:1987}. However, we will not
follow this approach for two reasons. First, the situation is more
subtle in the case of $\CL^{4|6}$ since $\CL^4$ only allows for a
nilpotent even direction inside $\CP^2\times \CP^2_*$ for
$\lambda_\pm=\mu_\pm$. Second, this description has several
drawbacks when the discussion of the Penrose-Ward transform
reaches the correspondence space, where the concepts of
thickenings (and the extended fattenings) are not sufficient, see
\cite{Eastwood:1987}.

\subsection{Third order thickenings and $d=4$ Yang-Mills theory}

Consider a vector bundle $E$ over the space $\FC^4\times \FC^4$
with coordinates $r^{\alpha\ald}$ and $s^{\alpha\ald}$. On $E$, we
assume a gauge potential $A=A^r_{\alpha\ald}\dd
r^{\alpha\ald}+A^s_{\beta\bed}\dd s^{\beta\bed}$. Furthermore, we
introduce the coordinates
\begin{equation}
x^{\alpha\ald}\ =\
\tfrac{1}{2}(r^{\alpha\ald}+s^{\alpha\ald})\eand k^{\alpha\ald}\
=\ \tfrac{1}{2}(r^{\alpha\ald}-s^{\alpha\ald})
\end{equation}
on the base of $E$. We claim that the Yang-Mills equations
$\nabla^{\alpha\ald} F_{\alpha\ald\beta\bed}=0$ are then
equivalent to
\begin{equation}\label{condYM4}
\begin{aligned}
{}[\nabla^r_{\alpha\ald},\nabla^r_{\beta\bed}]&\ =\
\ast[\nabla^r_{\alpha\ald},\nabla^r_{\beta\bed}]+\CO(k^3)~,\\
{}[\nabla^s_{\alpha\ald},\nabla^s_{\beta\bed}]&\ =\
-\ast[\nabla^s_{\alpha\ald},\nabla^s_{\beta\bed}]+\CO(k^3)~,\\
{}[\nabla^r_{\alpha\ald},\nabla^s_{\beta\bed}]&\ =\ \CO(k^3)~,
\end{aligned}
\end{equation}
where we define\footnote{One could also insert an $\di$ into this
definition but on $\FC^4$, this is not natural.} $\ast
F^{r,s}_{\alpha\ald\beta\bed}:=\frac{1}{2}\eps^{r,s}_{
\alpha\ald\beta\bed\gamma\gad\delta\ded}F_{r,s}^{\gamma\gad\delta\ded}$
separately on each $\FC^4$.

To understand this statement, note that equations \eqref{condYM4}
are equivalent to
\begin{equation}\label{condYM4simp}
\begin{aligned}
{}[\nabla^x_{\alpha\ald},\nabla^x_{\beta\bed}]&\ =\
[\nabla^k_{\alpha\ald},\nabla^k_{\beta\bed}]+\CO(k^3)~,\\
{}[\nabla^k_{\alpha\ald},\nabla^x_{\beta\bed}]&\ =\
\ast[\nabla^k_{\alpha\ald},\nabla^k_{\beta\bed}]+\CO(k^3)~,
\end{aligned}
\end{equation}
which is easily seen by performing the coordinate change from
$(r,s)$ to $(x,k$). These equations are solved by the expansion
\cite{Witten:1978xx,Isenberg:kk}
\begin{equation}\label{gaugepotYM4}
\begin{aligned}
A^k_{\alpha\ald}&\ =\
-\tfrac{1}{2}F^{x,0}_{\alpha\ald\beta\bed}k^{\beta\bed}-
\tfrac{1}{3}k^{\gamma\gad}\nabla^{x,0}_{\gamma\gad}(\ast
F^{x,0}_{\alpha\ald\beta\bed})k^{\beta\bed}~,\\
A^x_{\alpha\ald}&\ =\ A^{x,0}_{\alpha\ald} -\ast
F^{x,0}_{\alpha\ald\beta\bed}k^{\beta\bed}-
\tfrac{1}{2}k^{\gamma\gad}\nabla^{x,0}_{\gamma\gad}(
F^{x,0}_{\alpha\ald\beta\bed})k^{\beta\bed}~,
\end{aligned}
\end{equation}
if and only if
$\nabla_{x,0}^{\alpha\ald}F^{x,0}_{\alpha\ald\beta\bed}=0$ is
satisfied. Here, a superscript $0$ always denotes an object
evaluated at $k^{\alpha\ald}=0$. Thus we saw that a solution to
the Yang-Mills equations corresponds to a solution to equations
\eqref{condYM4} on $\FC^4\times \FC^4$.

As discussed before, the self-dual and anti-self-dual field
strengths solving the first and second equations of \eqref{condYM4}
can be mapped to certain holomorphic vector bundles over $\CP^3$
and $\CP^3_*$, respectively. On the other hand, the potentials
given in \eqref{gaugepotYM4} are now defined on a second order
infinitesimal neighborhood\footnote{not a thickening} of the
diagonal in $\FC^4\times\FC^4$ for which $\CO(k^3)=0$. In the
twistor description, this potential corresponds to a transition
function $f_{+-}\sim\psi_+^{-1}\psi_-$, where the \v{C}ech
0-cochain $\{\psi_\pm\}$ is a solution to the equations
\begin{equation}
\begin{aligned}
\lambda^\ald_\pm\left(\der{r^{\alpha\ald}}+A^r_{\alpha\ald}\right)\psi_\pm&\
=\ \CO(k^4)~,\\
\mu^\alpha_\pm\left(\der{s^{\alpha\ald}}+A^s_{\alpha\ald}\right)\psi_\pm&\
=\ \CO(k^4)~.
\end{aligned}
\end{equation}
Roughly speaking, since the gauge potentials are defined to order
$k^2$ and since $\der{r^{\alpha\ald}}$ and $\der{s^{\alpha\ald}}$
contain derivatives with respect to $k$, the above equations can
indeed be rendered exact to order $k^3$. The exact definition of
the transition function is given by
\begin{equation}
f_{+-,i}\ :=\
\sum_{j=0}^i\psi_{+,j}^{-1}\psi^{\phantom{-1}}_{-,i-j}~,
\end{equation}
where the additional indices label the order in $k$. On the
twistor space side, a third order neighborhood in $k$ corresponds
to a third order thickening in
\begin{equation}
\z{\kappa}_{(a)}\ :=\
z^\alpha_{(a)}\mu_\alpha^{(a)}-u^\ald_{(a)}\lambda_\ald^{(a)}~.
\end{equation}

Altogether, we see that a solution to the Yang-Mills equations
corresponds to a topologically trivial holomorphic vector bundle
over a third order thickening of $\CL^5$ in $\CP^3\times \CP^3_*$,
which becomes holomorphically trivial, when restricted to any
$\CPP^1\times \CPP^1_*\embd\CL^5$.

\subsection{Third order sub-thickenings and $d=3$ Yang-Mills-Higgs theory}

Let us now translate the above discussion to the three-dimensional
situation. First of all, the appropriate Yang-Mills-Higgs
equations obtained by dimensional reduction are
\begin{equation}\label{Yang-Mills-Higgs}
\nabla^{(\ald\bed)}F_{(\ald\bed)(\gad\ded)}\ =\
[\phi,\nabla_{(\gad\ded)}\phi]\eand\triangle \phi\ :=\
\nabla^{(\ald\bed)}\nabla_{(\ald\bed)}\phi\ =\ 0~,
\end{equation}
while the self-dual and anti-self-dual Yang-Mills equations
correspond after the dimensional reduction to two Bogomolny
equations which read as
\begin{equation}
F_{(\ald\bed)(\gad\ded)}\ =\
\eps_{(\ald\bed)(\gad\ded)(\dot{\eps}\dot{\zeta})}\nabla^{(\dot{\eps}\dot{\zeta}
)}\phi\eand
F_{(\ald\bed)(\gad\ded)}\ =\
-\eps_{(\ald\bed)(\gad\ded)(\dot{\eps}\dot{\zeta})}\nabla^{(\dot{\eps}\dot{\zeta
})}\phi~,
\end{equation}
respectively. Using the decomposition
$F_{(\ald\bed)(\gad\ded)}=\eps_{\ald\gad}f_{\bed\ded}+\eps_{\bed\ded}f_{\ald\gad
}$,
the above two equations can be simplified to
\begin{equation}
f_{\ald\bed}=\tfrac{\di}{2}\nabla_{(\ald\bed)}\phi\eand
f_{\ald\bed}=-\tfrac{\di}{2}\nabla_{(\ald\bed)}\phi~.
\end{equation}
Analogously to the four dimensional case, we start from a vector
bundle $E$ over the space $\FC^3\times \FC^3$ with coordinates
$p^{(\ald\bed)}$ and $q^{(\ald\bed)}$; additionally we introduce
the coordinates
\begin{equation}
y^{(\ald\bed)}\ =\
\tfrac{1}{2}(p^{(\ald\bed)}+q^{(\ald\bed)})\eand h^{(\ald\bed)}\
=\ \tfrac{1}{2}(p^{(\ald\bed)}-q^{(\ald\bed)})
\end{equation}
and a gauge potential
\begin{equation}
A=A^p_{(\ald\bed)}\dd p^{(\ald\bed)}+A^q_{(\ald\bed)}\dd
q^{(\ald\bed)}=A^y_{(\ald\bed)}\dd
y^{(\ald\bed)}+A^h_{(\ald\bed)}\dd h^{(\ald\bed)}
\end{equation}
on $E$. The differential operators we will consider in the
following are obtained from covariant derivatives by dimensional
reduction and take, e.g., the shape
\begin{equation}
\nabla_{\ald\bed}^y=\der{y^{(\ald\bed)}}+[A_{(\ald\bed)}^y+\tfrac{\di}{2}\eps_{
\ald\bed}\phi^y,\,\cdot\,]~.
\end{equation}
We now claim that the Yang-Mills-Higgs equations
\eqref{Yang-Mills-Higgs} are equivalent to the equations
\begin{equation}\label{condYM3}
\begin{aligned}
{}[\nabla^p_{\ald\bed},\nabla^p_{\gad\ded}]&\ =\
\ast[\nabla^p_{\ald\bed},\nabla^p_{\gad\ded}]+\CO(h^3)~,\\
{}[\nabla^q_{\ald\bed},\nabla^q_{\gad\ded}]&\ =\
-\ast[\nabla^q_{\ald\bed},\nabla^q_{\gad\ded}]+\CO(h^3)~,\\
{}[\nabla^p_{\ald\bed},\nabla^q_{\gad\ded}]&\ =\ \CO(h^3)~,
\end{aligned}
\end{equation}
where we can use
$\ast[\nabla^{p,q}_{\ald\bed},\nabla^{p,q}_{\gad\ded}]=
\eps_{\ald\bed\gad\ded\dot{\eps}\dot{\zeta}}\nabla_{p,q}^{\dot{\eps}\dot{\zeta}}
\phi^{p,q}$.
These equations can be simplified in the coordinates $(y,h)$ to
equations similar to \eqref{condYM4}, which are solved by the
field expansion
\begin{equation}\label{gaugepotYM3}
\begin{aligned}
A^h_{(\ald\bed)}&\ =\
-\tfrac{1}{2}F^{y,0}_{(\ald\bed)(\gad\ded)}h^{(\gad\ded)}-
\tfrac{1}{3}h^{(\gad\ded)}\nabla^{y,0}_{(\gad\ded)}
\eps_{(\ald\bed)(\dot{\eps}\dot{\zeta})(\dot{\sigma}\dot{\tau})}
(\nabla_{y,0}^{(\dot{\sigma}\dot{\tau})}\phi)h^{(\dot{\eps}\dot{\zeta})}~,\\
\phi^h&\ =\ \tfrac{1}{2}\nabla^{y,0}_{(\gad\ded)}\phi^{y,0}
h^{(\gad\ded)}+\tfrac{1}{6}h^{(\gad\ded)}\nabla^{y,0}_{(\gad\ded)}
\eps^{(\ald\bed)(\dot{\eps}\dot{\zeta})(\dot{\sigma}\dot{\tau})}
F^{y,0}_{(\dot{\eps}\dot{\zeta})(\dot{\sigma}\dot{\tau})}h_{(\ald\bed)}~,\\
A^y_{(\ald\bed)}&\ =\ A^{y,0}_{(\ald\bed)}
-\eps_{(\ald\bed)(\dot{\eps}\dot{\zeta})(\dot{\sigma}\dot{\tau})}
(\nabla_{y,0}^{(\dot{\sigma}\dot{\tau})}\phi^{y,0})h^{(\dot{\eps}\dot{\zeta})}-
\tfrac{1}{2}h^{(\gad\ded)}\nabla^{y,0}_{(\gad\ded)}(
F^{y,0}_{(\ald\bed)(\dot{\eps}\dot{\zeta})})h^{(\dot{\eps}\dot{\zeta})}~,\\
\phi^y&\ =\ \phi^{y,0}
+\tfrac{1}{2}\eps^{(\ald\bed)(\dot{\eps}\dot{\zeta})(\dot{\sigma}\dot{\tau})}
F^{y,0}_{(\dot{\eps}\dot{\zeta})(\dot{\sigma}\dot{\tau})}h_{(\ald\bed)}+
\tfrac{1}{2}h^{(\gad\ded)}\nabla^{y,0}_{(\gad\ded)}(\nabla_{(\ald\bed)}\phi^{y,0
}
)h^{(\ald\bed)}~,
\end{aligned}
\end{equation}
if and only if the Yang-Mills-Higgs equations
\eqref{Yang-Mills-Higgs} are satisfied.

Thus, solutions to the Yang-Mills-Higgs equations
\eqref{Yang-Mills-Higgs} correspond to solutions to equations
\eqref{condYM3} on $\FC^3\times \FC^3$. Recall that solutions to
the first two equations of \eqref{condYM3} correspond in the
twistor description to holomorphic vector bundles over
$\CP^2\times \CP^2_*$. Furthermore, the expansion of the gauge
potential \eqref{gaugepotYM3} is an expansion in a second order
infinitesimal neighborhood of $\diag(\FC^3\times \FC^3)$. As we
saw in the construction of the mini-superambitwistor space
$\CL^{4|6}$, the diagonal for which $h^{(\ald\bed)}=0$ corresponds
to $\CL^4\subset \CP^2\times \CP^2_*$. The neighborhoods of this
diagonal will then correspond to {\em sub-thickenings} of $\CL^4$
inside $\CP^2\times \CP^2_*$, i.e.\ for $\mu_\pm=\lambda_\pm$, we
have the additional nilpotent coordinate $\xi$. In other words,
the sub-thickening of $\CL^4$ in $\CP^2\times \CP^2_*$ is obtained
by turning one of the fiber coordinates of $\CP^2\times \CP^2$
over $\CPP^1_{\Delta}$ into a nilpotent even coordinate (in a
suitable basis). Then we can finally state the following:

Gauge equivalence classes of solutions to the three-dimensional
Yang-Mills-Higgs equations are in one-to-one correspondence with
gauge equivalence classes of holomorphic pseudo-bundles over a
third order sub-thickening of $\CL^4$, which become holomorphically trivial vector bundles when restricted to a $\CPP^1\times \CPP^1$ holomorphically embedded into $\CL^4$.

\section*{Acknowledgements}

I would like to thank Alexander Popov, Sebastian Uhlmann and
Martin Wolf for numerous helpful discussions and comments on a
draft of this paper. Furthermore, I am grateful to Klaus Hulek for
a discussion on the geometry of the mini-superambitwistor space
$\CL^{4|6}$. This work was done within the framework of the DFG
priority program (SPP 1096) in string theory.

\renewcommand{\baselinestretch}{0.9}

\end{document}